\documentclass[10pt]{sigplanconf}

\usepackage[utf8]{inputenc}
\usepackage[T1]{fontenc}
\usepackage{microtype}

\usepackage{algorithm2e}
\usepackage{amsmath}
\usepackage{booktabs}
\usepackage{cases}
\usepackage{enumerate}
\usepackage{enumitem}
\usepackage{graphicx}
\usepackage{listings}
\usepackage[hidelinks]{hyperref}
\usepackage{setspace}
\usepackage[]{caption}
\usepackage{subcaption}
\usepackage{tikz}
\usepackage{xcolor}
\usepackage{MnSymbol,wasysym}
\usepackage[firstpage]{draftwatermark}
\usepackage{flushend}

\SetWatermarkText{\hspace*{8in}\raisebox{3.9in}{\includegraphics[width=2.5cm]{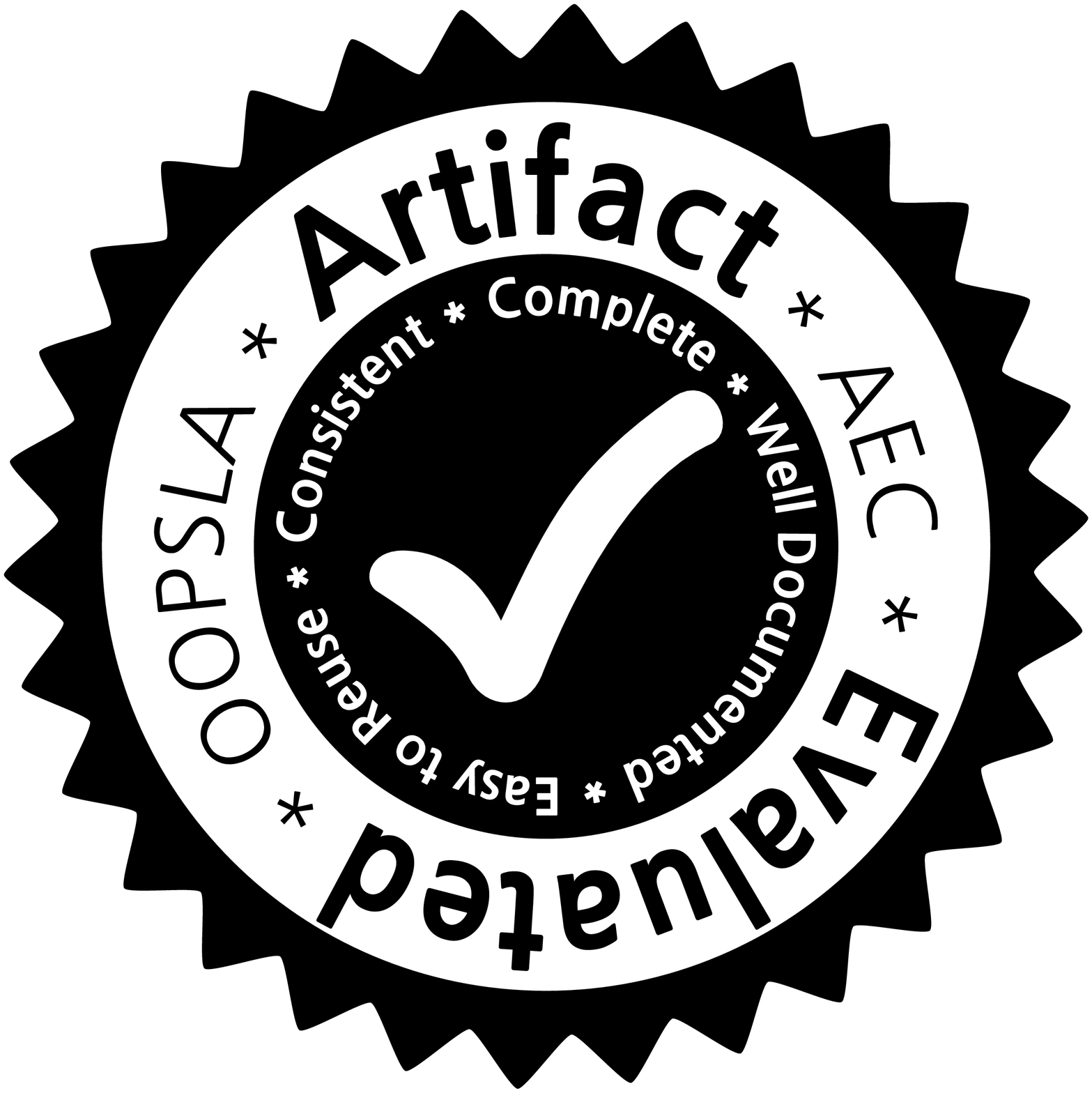}}}
\SetWatermarkAngle{0}

\definecolor{light-gray}{gray}{0.5}
\definecolor{dark-gray}{gray}{0.2}

\usetikzlibrary{%
  shapes,positioning,automata,arrows,decorations,patterns,%
  decorations.pathmorphing%
}

\newcommand{\OurSubsubsection}[1]{\smallbreak\noindent\textbf{#1}\xspace}

\newcommand{\CL}[1]{\label{code:#1}}
\newcommand{\lineRef}[1]{line~\ref{code:#1}}
\newcommand{\lineRangeRef}[2]{lines~\mbox{\ref{code:#1}--\ref{code:#2}}}

\newcommand{\impl}[1]{{\tt #1}\xspace}

\newcommand{\trans}[2]{{\it #1} $\rightarrow$ {\it #2}}

\let\oldbibitem\bibitem
\def\bibitem{\vfill\oldbibitem}

\newcommand{\ourSubcaption}[1]{%
  \vspace{-\medskipamount}%
  \caption{#1}%
  \vspace{-\medskipamount}%
}

\newcommand{\ourCaption}[1] {%
  \vspace{-\bigskipamount}%
  \caption{#1}%
  \vspace{\smallskipamount}%
}

\begin{document}

\setlength{\pdfpageheight}{\paperheight}
\setlength{\pdfpagewidth}{\paperwidth}

\conferenceinfo{OOPSLA'15}{October 25--30, 2015, Pittsburgh, PA, USA}
\copyrightyear{2015}
\copyrightdata{978-1-4503-3689-5/15/10}
\assigneddoi{2814270.2814294}

\permissiontopublish             %

\titlebanner{Preprint}        %

\title{Fast, Multicore-Scalable, Low-Fragmentation Memory Allocation
through Large Virtual Memory and Global Data Structures}

\authorinfo{Martin Aigner \and 
            Christoph M.~Kirsch \and
            Michael Lippautz \and
            Ana Sokolova}
           {University of Salzburg, Austria}
           {firstname.lastname@cs.uni-salzburg.at}

\maketitle

\begin{abstract}
We demonstrate that general-purpose memory allocation involving many threads on many cores can be done with high performance, multicore scalability, and low memory consumption.
For this purpose, we have designed and implemented scalloc, a concurrent allocator that generally performs and scales in our experiments better than
other allocators while using less memory, and is still competitive otherwise. The main ideas behind the design of scalloc are: uniform treatment of
small and big objects through so-called virtual spans, efficiently and
effectively reclaiming free memory through fast and scalable global data
structures, and constant-time (modulo synchronization) allocation and
deallocation operations that trade off memory reuse and spatial locality without being subject to false sharing.
\end{abstract}

\category{D.4.2}{Operating Systems}{Storage Management}[Allocation/deallocation strategies]
\category{D.3.4}{Programming Languages}{Processors}[Memory management (garbage collection)]

\keywords{Memory allocator, virtual memory, concurrent data structures, multicore scalability}

\section{Introduction}

Dynamic memory management is a key technology for high-level programming. Most of the existing memory allocators are extremely robust and well designed. Nevertheless, dynamic memory management for multicore machines is still a challenge. In particular, allocators either do not scale in performance for highly dynamic allocation scenarios, or they do scale but consume a lot of memory. Scalable performance of concurrent programs is crucial to utilize multicore hardware. Scalable allocation is needed for scalable performance of concurrent programs which allocate memory dynamically. 

Even though general purpose allocation was shown superior to custom allocation~\cite{Berger:OOPSLA02} a decade ago, many multi-threaded applications still use custom allocation or limit dynamic allocation to very few threads. We have developed a new memory allocator called scalloc that enables programmers to design many-threaded high-performance applications that dynamically allocate memory without the need for any additional application-tailored allocation strategies.

There are three high-level challenges in the design of a competitive concurrent allocator: (1) fast allocation for high performance, (2) memory layout facilitating fast access, and (3) effective and efficient reuse of memory for low memory consumption. All need to be achieved for various object sizes. Moreover, in presence of concurrency, all need to be achieved with as little coordination as possible. In particular, it is well-established common practice in concurrent allocators to allocate from thread-local allocation buffers (TLABs)~\cite{Berger:ASPLOS00} for fast allocation. The memory layout needs to support spatial locality without introducing false sharing. Finally, the key challenge is the reuse of free memory: When many threads deallocate objects of various sizes, even in other threads' TLABs, the freed memory should ideally be globally available for immediate reuse by any thread (requiring thread synchronization) for any request (requiring memory defragmentation). Coordination is necessary for memory reuse, the challenge is to minimize its need and cost.

Most competitive concurrent allocators are separated into a local (TLAB-based) frontend, for fast allocation, and a global backend, for reusing memory. TLABs grow and shrink incrementally in chunks of memory which we call spans. Note that spans are called superblocks in Hoard~\cite{Berger:ASPLOS00}, SLABs in llalloc~\cite{llalloc}, superpages in Streamflow~\cite{Schneider:ISMM06}, and spans in TCMalloc~\cite{tcmalloc}. Allocation is thread-local and fast as long as there is space in the allocating thread's TLAB, otherwise the TLAB needs to grow by a new span. Therefore, the larger the spans are, the faster the allocation. However, the larger the spans, the higher the memory consumption, since free memory in the TLAB of a given thread is not available to any other thread. Only when a span is empty it can be reused by other threads via the global backend, or even be returned to the operating system. The scalability of the global backend is crucial for the scalability of the allocator. Hence, there is a trade-off between scalability and memory consumption.

There are two possibilities in the design of spans: either all spans are of the same size, or not. If not all spans are of the same size, then making free spans reusable for as many requests as possible requires thread synchronization and defragmentation (to control external fragmentation). If they are of the same size, then free spans can be reused by all requests up to the size of the spans, only requiring thread synchronization. The disadvantage of same-sized spans that can accommodate small but also big objects is increased memory consumption (due to internal fragmentation). Most competitive concurrent allocators are therefore hybrids that use relatively small same-sized spans for small objects and different-sized spans for big objects.

To address the challenges in scalable concurrent allocation, we have designed scalloc based on three main ideas introduced here: (a) virtual spans which are same-sized spans in virtual memory; (b) a scalable global backend based on recently developed scalable concurrent data structures; and (c) a constant-time (modulo synchronization) frontend that eagerly returns spans to the backend. We have implemented these three new concepts in scalloc which is open source written in standard C/C++ and supports the POSIX API of memory allocators (malloc, posix\_memalign, free, etc.).

Scalloc provides scalability and reduces memory consumption at the same time. Virtual spans reduce the need and cost of coordination (as they are all of the same size) and at the same time reduce memory consumption. Namely, a virtual span contains a real span of a typically much smaller size for actual allocation. Due to on-demand paging, the rest of the virtual span does not manifest in actual memory consumption. The trick is that the operating system implicitly takes care of physical memory fragmentation because the fraction of the virtual span that is unused has no cost aside of consuming virtual address space. Moreover, the same size of virtual spans allows us to use a single global data structure for collecting and reusing empty spans. Global shared data structures were long considered performance and scalability bottlenecks that were avoided by introducing complex hierarchies. Recent developments in concurrent data structure design show that fast and scalable pools, queues, and stacks are possible~\cite{Henzinger:POPL13,Haas:CF13,Afek:OPODIS10}. We leverage these results by providing a fast and scalable backend, called the span-pool, that efficiently and effectively reuses and returns free memory to the operating system (through \impl{madvise} system calls). To the best of our knowledge, no other concurrent allocator uses a single global data structure as its backend. The frontend takes advantage of the scalable backend by eagerly returning spans as soon as they get empty. In contrast to other approaches, the frontend runs per-thread in constant time (modulo synchronization) meaning that at least one thread will make progress in constant time while others may have to retry. The combination of all three design choices is needed to achieve scalability and low memory consumption.

Scalloc is based on a mix of lock-free and lock-based concurrent data structures to minimize code complexity without sacrificing performance (scalloc is implemented in around 3000 lines of code). In our and others'~\cite{Hendler:SPAA10} experience locks are still a good choice for synchronizing data structures that are mostly uncontended.

Our experiments show that scalloc increases performance on average %
while consuming 
less memory than the previously fastest allocator (TBB).
Furthermore, scalloc outperforms and outscales all allocators for varying objects sizes that fit virtual spans while consuming as much or less memory. Scalloc handles spatial locality without being subject to active or passive false sharing, like some other allocators. Access to memory allocated by scalloc is as fast or faster than with most other allocators.

In the following section we provide high-level insight into the experience we gained when developing scalloc that is relevant to a broader audience beyond memory management experts. Section~\ref{sec:scalloc} explains the design of virtual spans, the span-pool, and the frontend and how they are integrated in scalloc. Section~\ref{sec:properties} discusses memory fragmentation and algorithmic complexity of scalloc. Section~\ref{sec:implementation} provides implementation details relevant for performance and reproducibility but optional for understanding the rest of the paper. We discuss related work in Section~\ref{sec:related-work} and present the experimental evaluation in Section~\ref{sec:experiments}.

\section{Experience and Relevance}

In our experience with multicore machines dynamic memory management may be a temporal and spatial bottleneck on machines with increasingly many cores and an increasing amount of memory. The challenge is to develop fast, scalable, and low-fragmentation allocators that provide fast memory access and are robust, i.e., do all of that for as many workloads as possible.

The following summary reflects our experience which we obtained in numerous experiments with different configurations and versions of scalloc as well as with many other state-of-the-art allocators.

\begin{itemize}
\item 64-bit address spaces and on-demand paging help reduce memory consumption and code complexity. On 64-bit machines even extremely large amounts of virtual address fragmentation can be tolerated because of on-demand paging and the sheer number of virtual addresses available. For example, when allocating 1-byte objects only, scalloc in its default configuration may still allocate up to 256GB of physical memory in 32TB of virtual memory (16KB real spans in 2MB virtual spans). As on-demand paging does not map unused virtual memory, defragmentation is only done (simply through unmapping) when resizing empty real spans. Virtual spans enable uniform treatment of a large range of object sizes (1-byte to 1MB objects in default scalloc) only leaving huge objects ($>$1MB) to the operating system. In particular, virtual spans enable the use of a single global data structure for collecting and reusing empty spans called the span-pool. However, that data structure still needs to scale in performance.

\item Lock-freedom improves performance and scalability but it may not always be necessary for overall performance and scalability. The span-pool is a so-called distributed stack~\cite{Haas:CF13} of lock-free Treiber stacks~\cite{Treiber86} (stacks rather than queues for spatial locality). In default scalloc there are as many Treiber stacks in the span-pool as there are cores on the machine. Access to the span-pool works by first identifying one of the Treiber stacks, which is fast, and then operating on that, possibly in parallel with as many threads as there are cores. The key insight here is that a span-pool with only one Treiber stack does not scale anymore but, most interestingly, results in slightly lower memory consumption because threads find empty spans even faster. Moreover, replacing Treiber stacks with lock-based stacks results in loss of performance and limits scalability~\cite{Henzinger:POPL13}. In a different part of scalloc we nevertheless use locks to synchronize access to an uncontended double-ended queue. Here, locking significantly reduces code complexity while not harming overall performance and scalability.

\item Constant-time deallocation is not only possible and obviously good for robustness but can also be done to save memory. Allocating objects thread-locally without synchronization is fast and standard practice. Deallocating objects, however, is more difficult as it may happen concurrently in the same span. Encouraged by the performance and scalability of the span-pool, a technical innovation in scalloc is that spans are eagerly inserted into the span-pool as soon as they get empty, replacing possibly non-constant-time cleanup later. It turns out that doing so reduces memory consumption further, again without harming performance and scalability.

\end{itemize}

On top of the design choices, a careful implementation of all concepts is necessary for competitive performance. We note that when comparing with other allocators, the implementations (and not necessarily the design choices) are being compared. Even a great concept may not perform when poorly implemented.   

\section{Virtual Spans, Span-Pool, and Frontend}\label{sec:scalloc}

This section explains on conceptual level how virtual spans, the span-pool, and the frontend work, and how they are integrated in scalloc.

\subsection{Real Spans and Size Classes}

Like many other allocators, scalloc uses the well-known concept of size classes.
A (real) span in scalloc is a contiguous portion of memory partitioned into same-sized
blocks. The size of blocks in a span determines the size class of the span. All spans in a given size class have the same number of blocks. In scalloc, there are 29 size classes but only 9 distinct real-span
sizes which are all multiples of 4KB (the size of a system page).

The first 16 size classes, with block sizes ranging from 16 bytes to 256 bytes
in increments of 16 bytes, are taken from TCMalloc~\cite{tcmalloc}.  These 16 size classes all have the same real-span
size. Size classes for larger blocks range from 512 bytes to 1MB, in increments
that are powers of two. These
size classes may have different real-span size, explaining the difference
between 29 size classes and 9 distinct real-span sizes. The design
of size-classes limits block internal fragmentation for sizes larger than
16 bytes to less than 50\%. 
  
Objects of size larger than any size class are not managed by spans, but rather
allocated directly from the operating system using {\tt mmap}.

\subsection{Virtual Spans}

\begin{figure}
\centering
\begin{tikzpicture}[scale=0.8, every node/.style={scale=0.8}]

\fill[black!5] (0,0) rectangle (2,4);
\draw[-,decorate,decoration={snake,amplitude=.4mm,segment length=1.5mm,post length=0mm}] (0,3.1) -- (0,3.6);
\draw[-,decorate,decoration={snake,amplitude=.4mm,segment length=1.5mm,post length=0mm}] (2,3.1) -- (2,3.6);
\draw[-] (0,3.6) -- (0,4) -- (2,4) -- (2,3.6);
\draw[-] (0,2) -- (0,3.1);
\draw[-] (2,2) -- (2,3.1);

\node[text=black!50,rotate=45,align=center] at (1,3) {\small reserved\\ \small addresses};

\draw[fill=blue!40] (0,0) rectangle (2,.5);
\draw[fill=blue!40] (0,.5) rectangle (2,1);
\draw[fill=blue!40] (0,1) rectangle (2,1.5);
\draw[fill=blue!40] (0,1.5) rectangle (2,2);

\node[] at (1, .75) {\scriptsize virtual span};

\fill[black!5] (3,0) rectangle (5,4);
\draw[-,decorate,decoration={snake,amplitude=.4mm,segment length=1.5mm,post length=0mm}] (3,3.1) -- (3,3.6);
\draw[-,decorate,decoration={snake,amplitude=.4mm,segment length=1.5mm,post length=0mm}] (5,3.1) -- (5,3.6);
\draw[-] (3,3.6) -- (3,4) -- (5,4) -- (5,3.6);
\draw[-] (3,1.5) -- (3,3.1);
\draw[-] (5,1.5) -- (5,3.1);
\node[text=black!50,rotate=45,align=center] at (4,3) {\small unmapped\\ \small memory};

\draw[fill=yellow!40] (3,0) rectangle (5,1.5);
\node[] at (4, .75) {\scriptsize real span};

\draw[fill=red!15] (6,0) rectangle (8,4);
\draw[fill=black!20] (6,0) rectangle (8,.5);
\node at (7,.25) {\scriptsize header};

\draw[pattern=north west lines, pattern color=green!15] (6,.5) rectangle (8,4);

\node[align=center] at (7,2.25) {\scriptsize block\\ \scriptsize payload};

\node at (1, 4.25) {arena};
\node at (4, 4.25) {virtual span};
\node at (7, 4.25) {real span};

\draw[dashed] (2, .5) -- (3,0);
\draw[dashed] (2, 1) -- (3, 4);

\draw[dashed] (5,0) -- (6,0);
\draw[dashed] (5,1.5) -- (6,4);

\end{tikzpicture}
\caption{Structure of arena, virtual spans, and real spans}
\label{fig:virtuals-spans}
\end{figure}
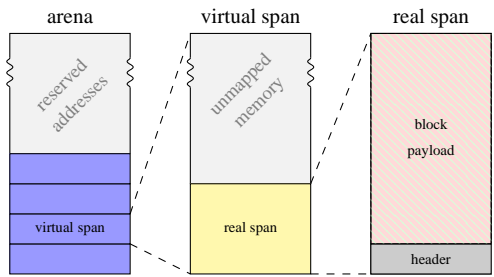

A virtual span is a span allocated in a very large portion of virtual memory
(32TB) which we call \emph{arena}. All virtual spans have the same fixed size of
2MB and are 2MB-aligned in the arena. Each virtual span contains a real span, of
one of the available size classes. Wa say ``size class of the virtual span'' for
the size class of the contained real span. Typically, the real span is (much)
smaller than the virtual span that contains it. The maximal real-span size is
limited by the size of the virtual span. This is why virtual spans are suitable
for big objects as well as for small ones. The structure of the arena, virtual
spans, and real spans is shown in Figure~\ref{fig:virtuals-spans}. The
advantages of using virtual spans are:

\begin{enumerate}[noitemsep]
  \item Virtual memory outside of real spans does not cause
    fragmentation of physical memory, as it is not used and therefore
    not mapped (because of on-demand paging of the operating system);
  \item Uniform treatment of small and big objects;
  \item No repeated system calls upon every span allocation since the arena
    is mmapped only once. 
\end{enumerate}

Note that since virtual spans are of the same size and aligned in virtual
memory, getting a new virtual span from the arena is simply incrementing a bump
pointer. When a virtual span gets empty, it is inserted into the free-list of
virtual spans, i.e., the span-pool discussed in the next section. The
disadvantages of using virtual spans are:
\begin{enumerate}[noitemsep]
  \item Current kernels and hardware only provide a 48-bit, instead of a 64-bit,
    address space.  As a result, not all of virtual memory can be utilized (see
    below);
  \item Returning a virtual span to the span-pool may be costly in one scenario: 
    a virtual span with a real span of a given size greater than a given
    threshold becomes empty and is inserted into the span pool. Then, in order
    to limit physical-memory fragmentation, we use madvise\footnote{
    \impl{madvise} is used to inform the kernel that a a range of
    virtual memory is not needed and the corresponding page frames can
    be unmapped.} to inform the kernel that the remaining virtual (and thus
    mapped physical) memory is no longer needed.
\end{enumerate}

Note that the design of the span-pool minimizes the chances
that a virtual span changes its real-span size.

Mmapping virtual memory in a single call at this order of
magnitude (32TB) is a new idea first developed for scalloc. Upon initialization,
scalloc mmaps $2^{45}$ virtual memory addresses, the upper limit for a single
mmap call on Linux. This call does not introduce any significant overhead as the
memory is not mapped by the operating system. It is still possible to allocate
additional virtual memory using mmap, e.g. for other memory allocation or
memory-mapped I/O.  The virtual address space still left is $2^{48} - 2^{45}$
bytes, i.e., 224TB.

In the worst case of the current configuration with 2MB virtual spans, if real
spans are the smallest possible (16KB), the physical memory addressable with
scalloc is
$(2^{45}/ 2^{21})\cdot 2^{14}
\text{~bytes~} = 2^{38} \text{~bytes~} = 256\rm{GB}$.

We have also experimented with configurations of up to 128MB for virtual spans
resulting in unchanged temporal and spatial performance for the benchmarks that
were not running out of arena space. Enhancing the Linux kernel to support
larger arenas is future work. On current hardware, with up to 48 bits for
virtual addresses, this would enable up to 256TB arena space and 2TB addressable
physical memory (in the worst case, with 2MB virtual spans and 16KB real spans).

Note that in scalloc virtual spans do not restrict the possibility of observing
segmentation faults because unmapped memory that is not used by a real-span is
still protected against access using the \impl{mprotect} system call.

\subsection{Backend: Span-Pool}

\begin{figure}[t]
\centering
\begin{tikzpicture}[scale=0.8, every node/.style={scale=0.8}]

\newcommand{\colHeight}{.8}
\newcommand{\halfColHeight}{.4}

\tikzstyle{common}=[
  align=center,
  execute at begin node=\setlength{\baselineskip}{.8em}
]

\draw (0,0) rectangle (2,\colHeight);
\node[common] at (1,\halfColHeight) {\scriptsize real-span size $x$};
\draw (2,0) rectangle (4,\colHeight);
\node[common] at (3,\halfColHeight) {\scriptsize real-span size $y$};
\draw (5,0) rectangle (7,\colHeight);
\node[common] at (6,\halfColHeight) {\scriptsize real-span size $n$};

\draw [-,decorate,decoration={snake,amplitude=.4mm,segment length=2mm,post length=0mm}] (4,{0+\colHeight}) -- (5,{0+\colHeight});
\draw [-,decorate,decoration={snake,amplitude=.4mm,segment length=2mm,post length=0mm}] (4,0) -- (5,0);

\draw (0,-1.2) rectangle (2,{-1.2+\colHeight});
\node[common] at (1,{-1.2+\halfColHeight}) {\scriptsize stack 1};
\draw (2,-1.2) rectangle (4,{-1.2+\colHeight});
\node[common] at (3,{-1.2+\halfColHeight}) {\scriptsize stack 2};
\draw (5,-1.2) rectangle (7,{-1.2+\colHeight});
\node[common] at (6,{-1.2+\halfColHeight}) {\scriptsize stack $p$};

\draw [-,decorate,decoration={snake,amplitude=.4mm,segment length=2mm,post length=0mm}] (4,{-1.2+\colHeight}) -- (5,{-1.2+\colHeight});
\draw [-,decorate,decoration={snake,amplitude=.4mm,segment length=2mm,post length=0mm}] (4,-1.2) -- (5,-1.2);


\draw[dashed] (2,0) -- (0,{-1.2+\colHeight});
\draw[dashed] (4,0) -- (7,{-1.2+\colHeight});

\draw[->] (3, -1.2) -- (3, -1.6);
\draw[->] (6, -1.2) -- (6, -1.6);
\draw[->] (1, -1.2) -- (1, -1.6);

\draw[fill=blue!40] (0.3,-1.6) rectangle (1.7,{-1.6-.4});
\node[common] at (1,{-1.6-.2}) {\scriptsize free  \scriptsize span};
\draw[->] (1.7,-1.7) -- (1.9,-1.7) -- (1.9,-2) -- (1.7,-2);
\draw[fill=blue!40] (0.3,-1.6-.4) rectangle (1.7,{-1.6-.8});
\node[common] at (1,{-1.6-.6}) {\scriptsize free  \scriptsize span};
\draw[->] (1.7,-2.1) -- (1.9,-2.1) -- (1.9,-2.4) -- (1.7,-2.4);
\draw[fill=blue!40] (0.3,-1.6-.8) rectangle (1.7,{-1.6-1.2});
\node[common] at (1,{-1.6-1}) {\scriptsize free  \scriptsize span};
\draw[->] (1.7,-2.5) -- (1.9,-2.5) -- (1.9,-2.7);
\node at (1.9, -2.8) {\scriptsize $\bot$};

\draw[fill=blue!40] (2.3,-1.6) rectangle (3.7,{-1.6-.4});
\node[common] at (3,{-1.6-.2}) {\scriptsize free  \scriptsize span};
\draw[->] (3.7,-1.7) -- (3.9,-1.7) -- (3.9,-2) -- (3.7,-2);
\draw[fill=blue!40] (2.3,-1.6-.4) rectangle (3.7,{-1.6-.8});
\node[common] at (3,{-1.6-.6}) {\scriptsize free  \scriptsize span};
\draw[->] (3.7,-2.1) -- (3.9,-2.1) -- (3.9,-2.4) -- (3.7,-2.4);
\draw[fill=blue!40] (2.3,-1.6-.8) rectangle (3.7,{-1.6-1.2});
\node[common] at (3,{-1.6-1}) {\scriptsize free  \scriptsize span};
\draw[->] (3.7,-2.5) -- (3.9,-2.5) -- (3.9,-2.7);
\node at (3.9, -2.8) {\scriptsize $\bot$};

\draw[fill=blue!40] (5.3,-1.6) rectangle (6.7,{-1.6-.4});
\node[common] at (6,{-1.6-.2}) {\scriptsize free  \scriptsize span};
\draw[->] (6.7,-1.7) -- (6.9,-1.7) -- (6.9,-2) -- (6.7,-2);
\draw[fill=blue!40] (5.3,-1.6-.4) rectangle (6.7,{-1.6-.8});
\node[common] at (6,{-1.6-.6}) {\scriptsize free  \scriptsize span};
\draw[->] (6.7,-2.1) -- (6.9,-2.1) -- (6.9,-2.4) -- (6.7,-2.4);
\draw[fill=blue!40] (5.3,-1.6-.8) rectangle (6.7,{-1.6-1.2});
\node[common] at (6,{-1.6-1}) {\scriptsize free  \scriptsize span};
\draw[->] (6.7,-2.5) -- (6.9,-2.5) -- (6.9,-2.7);
\node at (6.9, -2.8) {\scriptsize $\bot$};

\draw[<->] (7.2, -1.2) -- (7.2,\colHeight);
\node[rotate=90] at (7.45,-.2) {\scriptsize pre allocated};

\draw[<->] (7.2, -1.6) -- (7.2,{-2.8});
\node[rotate=90] at (7.45,-2.2) {\scriptsize dynamic};

\end{tikzpicture}
\caption{Span-pool layout }
\label{fig:span-pool}
\end{figure}
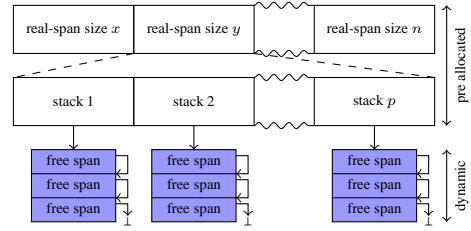

The span-pool is a global concurrent data structure that logically corresponds
to an array of real-span-size-segregated ``stack-like'' pools. The span-pool
implements put and get methods; no values are lost nor invented from thin air;
it neither provides a linearizable emptiness check, nor any order
guarantees. However, each pool within the span-pool is a locally
linearizable~\cite{Haas:LocLin15} ``stack-like'' pool. It is ``stack-like''
since in a single-threaded scenario it is actually a stack.

Figure~\ref{fig:span-pool} illustrates that the segregation by real-span size is
implemented as pre-allocated array where each index in the array refers to a
given real-span size. Consequently, all size classes that have the same
real-span size refer to the same index. Each array entry then holds another
pre-allocated array, the pool array, this time of lock-free Treiber
stacks~\cite{Treiber86}. The pool array has size equal to the number of cores
(determined at run-time during the initialization phase of the allocator). As a
result a stack in any of the pools of the span-pool is identified by a real-span
index and a core index. A Treiber stack is a pointer to the top of a singly-linked list of elements; pushing and popping is done lock-free by atomic compare-and-swap operations on the top pointer. 

The design is inspired by distributed queues~\cite{Haas:CF13}.  We use stacks
rather than queues for the following reasons: spatial locality, especially on
thread-local workloads; lower latency of push and pop compared to enqueue and
dequeue; and stacks can be implemented without sentinel nodes, i.e., no
additional memory is needed for the data structure. Thereby, we utilize the
memory of the elements inserted into the pool to construct the stacks, avoiding
any dynamic allocation of administrative data structures. Distributed stacks
are, to our knowledge, among the fastest scalable pools. To make the occurrence
of the ABA problem~\cite{Herlihy08} unlikely we use 16-bit ABA counters that are
embedded into link pointers\footnote{Currently a 64-bit address space is
limited to 48 bits of address, enabling the other 16 bits to be used as ABA
counter.}. Completely avoiding the ABA problem is a non-trivial task, which can
be solved using e.g. hazard pointers~\cite{Michael:IEEE04}.

\begin{figure}[t]
\begin{lstlisting}[
    caption={Span-pool pseudo code},
    label={lst:span-pool},  xleftmargin=3.0ex]
Int num_cores();  // Returns the number of cores.
Int thread_id();  // Returns this thread's id 
                  // (0-based).

// Utility functions to map spans and real-span
// sizes to distinct indexes.
Int real_span_idx(Span span);
Int real_span_idx(Int real_span_size);

// Returns the real span size of a given span.
Int real_span_size(Span span);

// Madvise all but a span's first page with
// MADV_DONTNEED.
void madvise_span(Span span);

SpanPool {
  Stack spans[MAX_REAL_SPAN_IDX][num_cores()];

  void put(Span span):
    Int rs_idx = real_span_idx(span);         (*\CL{put_rs_idx} *)
    Int core_idx = thread_id() mod num_cores(); (*\CL{put_core_idx} *)
    if real_span_size(span) >= MADVISE_THRESHOLD:(*\CL{put_madvise_th}*)
      madvise_span(span);                     (*\CL{put_madvise}*)
    spans[rs_idx][core_idx].put(span);(*\CL{put_stack}*)

  Span get(Int size_class):
    Int rs_idx = real_span_idx(size_class);   (*\CL{get_rs_idx}*)
    Int core_idx = thread_id() mod num_cores();(*\CL{get_core_idx}*)
    // Fast path.
    spans[rs_idx][core_idx].get();           (*\CL{get_fast_path}*)
    if span == NULL:
      // Try to reuse some other span.
      for rs_idx in range(0, MAX_REAL_SPAN_IDX): (*\CL{get_search_start}*)
        for core_idx in range(0, num_cores()):
          spans_[rs_idx][core_idx].Get();
          if span != NULL:
            return span;                      (*\CL{get_search_end}*)
    // If everything fails, just return a span from
    // the arena.
    return arena.allocate_virtual_span();     (*\CL{get_arena_call}*)
}
\end{lstlisting}
\end{figure}

Listing~\ref{lst:span-pool} shows the pseudo code of the span-pool. Upon
returning a span to the span-pool, a thread performing a \texttt{put} call first
determines the real-span index for a given span~(\lineRef{put_rs_idx}) and the
core index as thread identifier modulo the number of
cores~(\lineRef{put_core_idx}). Before actually inserting~(\lineRef{put_stack})
the given span into the corresponding stack the thread may return the span's
underlying memory to the operating system using the {\tt  madvise} system call
with advice {\tt MADV\_DONTNEED}~(\lineRef{put_madvise}), effectively freeing
the affected memory. This is the expensive case, only performed on spans with
large real-span size determined by a threshold, as unused spans with large
physically mapped real-spans result in noticeable physical fragmentation and the
{\tt madvise} system call may anyway be necessary upon later reuse, e.g. when a
span is to be reused in a size class with a smaller real-span size. The 
\impl{MADVISE\_THRESHOLD}~(\lineRef{put_madvise_th}) is set to 32KB, which is
the boundary between real-span sizes of size  classes that are incremented by 16
bytes and those that are incremented in powers of two. Note that lowering the
threshold does not substantially improve the observed memory consumption in our
experiments while it noticeably decreases performance. Furthermore, for
scenarios where physical fragmentation is an issue, one can add a compaction
call that traverses and madvises particular spans.

Upon retrieving a span from the span pool, for a given size class, a  thread
performing a \texttt{get} call first determines the real-span index of the size
class~(\lineRef{get_rs_idx}) and the core index as thread identifier modulo
the number of cores~(\lineRef{get_core_idx}). In the fast path for span
retrieval the thread then tries to retrieve a span from this identified
stack~(\lineRef{get_fast_path}). Note that this fast path implements the
match to the \texttt{put} call, effectively maximizing locality for consecutively
inserted) and retrieved spans of equal real-span sizes. If no span is
found in the fast path, the thread searches all real-span size indexes and core
indexes for a span to
use~(\lineRangeRef{get_search_start}{get_search_end}). Note that this
motivates the design of the real-span sizes: For reuse, a span of a large
real-span size has anyway been madvised whereas all other spans have the same
real-span size; Reusing a span in the same real-span size (even if the size
class changes) amounts only to changing the header. Only when the search for an
empty virtual span  fails, the thread gets a new virtual span from the arena (as
for initial allocation; \lineRef{get_arena_call}). Note that the search
through the span-pool may fail even if there are spans in it due to the global
use of the arrays (and the nonlinearizable emptiness check).

\subsection{Frontend: Allocation and Deallocation}

We now explain the mutator-facing frontend of scalloc, i.e., the part of the
allocator that handles allocation and deallocation requests from the mutator.

\begin{figure}[t]
\centering
\begin{tikzpicture}[node distance=1cm,shorten >=2pt,scale=0.8, every node/.style={scale=0.8}]




\tikzstyle{common}=[
fill=black, circle, inner sep=1.5pt
]

\node[common] (free) {};
\node[] at(free) [xshift=14pt,yshift=10pt] {\small free};
\node[common] (hot) [below left=of free] {};
\node[] at(hot) [xshift=-13pt] {\small hot};
\node[common] (cold) [below right=of hot] {};
\node[] at(cold) [yshift=-10pt] {\small floating};
\node[common] (full) [below right=of free] {};
\node[] at(full) [xshift=20pt] {\small reusable};
\node[common,initial,initial text={}] (dead) [above=of free,yshift=-.5cm] {};
\node[] at(dead) [yshift=10pt] {\small expected};

\path[->] (free) edge[bend right,double] 
  (hot);
\path[->] (hot)  edge[bend right,double] 
  (cold);
\path[->] (cold) edge[bend right] 
  (full);
\path[->] (full) edge[bend right]
  (free);
\path[->] (dead) edge[double]
  (free);
\path[->] (full) edge[double]
  (hot);

\draw[-,dashed,decorate,decoration={snake,amplitude=.2mm,segment length=4mm,post length=0mm}] (-2,.7) -- (3,.7);

\draw[-,dashed,decorate,decoration={snake,amplitude=.2mm,segment length=4mm,post length=0mm}] (-2,-.3) -- (3,-.3);

\node[text width=4cm] at (4.5,1) {\scriptsize Arena (RSS $= 0$)};
\node[text width=4cm] at (4.5,.2) {\scriptsize Backend (RSS compacted)};
\node[text width=4cm] at (4.5,-.6) {\scriptsize Frontend (RSS $=$ real span size)};

\draw[-] (2.5, -2) edge[double] (3, -2);
\node[text width=4cm] at (5, -2) {\small malloc()};
\draw[-] (2.5, -2.35) edge[] (3, -2.35);
\node[text width=4cm] at (5, -2.35) {\small free()};

\end{tikzpicture}
\caption{Life cycle of a span}
\label{fig:span-life}
\end{figure}
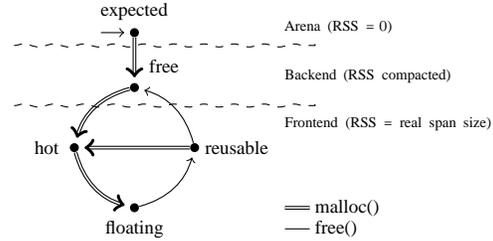

We distinguish several states in which a span can be, illustrated in
Figure~\ref{fig:span-life}. A span can be in a state: \emph{expected},
\emph{free}, \emph{hot}, \emph{floating}, or \emph{reusable}. A span is 
\emph{expected} if it is still in the arena, i.e., it is completely unused. Note
that in this state its memory footprint is 0 bytes. Spans contained in the
span-pool are \emph{free}. A span can be in some of the other states only when
it is in the frontend, i.e., it is assigned a specific size class. Spans that
are \emph{hot} are used for allocating new blocks. For spans that are not hot we
distinguish between floating and reusable based on a threshold of the number of
free blocks. Spans with less than or equal free blocks than the specified
threshold are \emph{floating}, spans with more free blocks than specified by the
threshold are \emph{reusable}. We refer to this threshold as \emph{reusability
threshold}. It is possible to only have spans that are floating, i.e., no reuse
of nonempty spans, at the expense of increased fragmentation. Throughout its
life in the frontend, a span is always assigned to exactly one local-allocation buffer (LAB), the so-called owning LAB. By default LABs are TLABs in scalloc, i.e., each LAB has a owner thread. Alternatively, scalloc can also be configured to use core-local allocation buffers (CLABs), i.e., one LAB per core where threads with equal identifiers modulo number of cores share the same LAB. Either way, in each LAB and for each size class there is a unique hot span. Furthermore, each LAB contains for each size class a set of reusable spans. More details about this set are given in Section~\ref{sec:implementation}.

A consequence of the concept
of ownership is that deallocation of a block may happen in spans that are not
owned by a thread. We refer to such deallocation as remote free, whereas
deallocation in a span owned by a thread is a local free. All allocations in
scalloc are done locally, performed in the corresponding hot span. A common problem in
allocator design is handling remote frees in a scalable way. Having no mechanism for handling remote frees results in so-called blowup
fragmentation~\cite{Berger:ASPLOS00}, i.e., any memory freed through remote frees cannot be reused again. Similar to other span-based
allocators~\cite{Schneider:ISMM06,llalloc}, scalloc provides two free lists of
blocks in each span, a local free list and a remote free list. The local free
list is only accessed by an owning thread, while the remote free list can be
accessed concurrently by multiple (not owning) threads at the same time.

\OurSubsubsection{Allocation.}
Upon allocation of a block in a given size class, a thread checks its LAB's size
class for a hot span. If a hot span exists the thread tries to find a block in
the local free list of the hot span. If a block is found, the thread allocates
in this block (this is the allocation fast path). The following situations can
also occur~(for implementation details see Section~\ref{sec:implementation}):
\begin{enumerate}[label=(\alph*),noitemsep]
  \item \label{lbl:no-span} No hot span exists in the given size class. The
    thread then tries to assign a new hot span by trying to reuse one from the
    set of reusable spans. If no span is found there, the thread falls back to
    retrieving a span from the span-pool.
  \item There is a hot span, but its local free list is empty. The idea now is
    to use a remotely freed block. However, it is not a wise option to allocate
    in the remote free list, as that would make allocation interfere with remote
    frees, destroying the performance of allocation. Therefore, this is a point
    of choice: If there are enough blocks (in terms of the reusability
    threshold) in the remote free list, the thread moves them all to its local
    free list and continues with fast-path allocation. Otherwise, if there are
    not enough blocks in the remote free list, the thread gets a new hot span
    like in~\ref{lbl:no-span}.
\end{enumerate}

\OurSubsubsection{Deallocation.}
Upon deallocation, a thread returns the block to be freed to the corresponding
free list, which is the local free list in case where the thread owns the span, and
the remote free list otherwise. Depending on the state of the span where the
block is allocated, the thread then performs the following actions~(for
implementation details see Section~\ref{sec:implementation}):
\begin{enumerate}[label=(\alph*),noitemsep]
  \item The span is floating. If the number of free blocks in this span is now
    larger than the reusability threshold, the span's state changes to reusable
    and the span is inserted into the set of reusable spans of the owning LAB.
  \item The span is reusable. If the free was the last free in the span, i.e.,
    all blocks have been freed, the span is removed from the set of reusable
    spans of the owning LAB and returned to the span-pool.
\end{enumerate}
If the span is hot, no additional action is taken.

Note that a new contribution of scalloc is that a span is freed upon the
deallocation of the last object in the span. All other span-based allocators
postpone freeing of a span until the next allocation which triggers a cleanup.

\section{Properties} \label{sec:properties}

\OurSubsubsection{Span-internal Fragmentation.}
Span-internal fragmentation is a global property and refers to memory assigned
to a real-span of a given size class that is currently free (unused by the
mutator) but cannot be reused for serving allocation requests in other size
classes by any LAB.

Let $f$ be the current global span-internal fragmentation. Let $s$ refer to the
span on which the next operation happens. Let $size$ be the size class of $s$
and $u$ be the size of the payload (memory usable for blocks) of $s$. At
initialization $f = 0$.

Then, for an allocation of a block in $s$

\scriptsize
\begin{numcases}{f=}
f + u - size & if no usable span~\label{eq:no-span}\\
f - size & otherwise~\label{eq:span}
\end{numcases}
\normalsize
where no usable spans means no hot span and no reusable spans are
present~\eqref{eq:no-span}. Note that a span might already be reusable (with
respect to
the threshold) but not yet present in the set of reusable spans. This case is
still covered by \eqref{eq:no-span} and is a result of the fact that freeing an
object and further processing it (reusable sets, or span-pool) are operations
performed non-atomically.

Furthermore, for a deallocation of a block in $s$

\scriptsize
\begin{numcases}{f=}
f - u & if last block~\label{eq:free-span}\\
f + size & otherwise~\label{eq:regular-free}
\end{numcases}
\normalsize
where last block~\eqref{eq:free-span} refers to the last free of a block in a
given span. Note that to achieve this fragmentation property on a free call an
allocator, such as scalloc, has to return an empty span to a global backend
immediately. A regular free not emptying the span increases fragmentation by the
size of the block as this span cannot be reused
globally~\eqref{eq:regular-free}.

\OurSubsubsection{Operation Complexity.}
An allocation operation only considers hot spans and reusable
spans and does not need to clean up empty spans. The operation is constant-time
as in the uncontended case either a hot span is present and can be used for
allocating a block or a reusable span is made hot again before
allocating a block in it. In the contended case more than one reusable
span may need to be considered because of concurrent deallocation operations.
At least one of the operations will make progress
in constant time.

A deallocation operation only considers the affected span, i.e.,
the span containing the block that is freed.
Local deallocations are constant-time and
remote deallocations are constant-time modulo synchronization
(insertion into the remote free-list which is lock-free).
Spans that get reusable are made reusable
in constant-time modulo synchronization
(insertion into the set of reusable spans which is lock-based).
Spans that get empty are handled in the span-pool.

Span-pool put and get operations are constant-time modulo synchronization
(the span-pool is lock-free).

\section{Implementation Details}\label{sec:implementation}

We now explain the implementation details of scalloc, i.e., the encoding of
fields in headers and the concrete algorithms used for allocation, deallocation,
and thread termination.

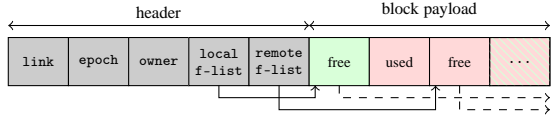
\begin{figure}[t]
\centering
\begin{tikzpicture}[scale=0.8, every node/.style={scale=0.8}]

\newcommand{\colHeight}{.8}
\newcommand{\halfColHeight}{.4}

\draw[fill=black!20] (0,0) rectangle (1,\colHeight);
\node at (.5,{.5*\colHeight}) {\scriptsize\tt link};
\draw[fill=black!20] (1,0) rectangle (2,\colHeight);
\node at (1.5,{.5*\colHeight}) {\scriptsize\tt epoch};
\draw[fill=black!20] (2,0) rectangle (3,\colHeight);
\node at (2.5,{.5*\colHeight}) {\scriptsize\tt owner};
\draw[fill=black!20] (3,0) rectangle (4,\colHeight);
\node[align=center,execute at begin node=\setlength{\baselineskip}{.8em}] at (3.5,{.5*\colHeight}) {\scriptsize\tt local\\ \scriptsize\tt f-list};
\draw[fill=black!20] (4,0) rectangle (5,\colHeight);
\node[align=center,execute at begin node=\setlength{\baselineskip}{.8em}] at (4.5,{.5*\colHeight}) {\scriptsize\tt remote\\ \scriptsize\tt f-list};

\draw[fill=green!15] (5,0) rectangle (6,\colHeight);
\node at (5.5,\halfColHeight) {\scriptsize free};

\draw[fill=red!15] (6,0) rectangle (7,\colHeight);
\node at (6.5,\halfColHeight) {\scriptsize used};

\draw[fill=red!15] (7,0) rectangle (8,\colHeight);
\node at (7.5,\halfColHeight) {\scriptsize free};

\draw[fill=red!15] (8,0) rectangle (9,\colHeight);
\draw[pattern=north west lines, pattern color=green!15] (8,0) rectangle (9,\colHeight);
\node at (8.5,\halfColHeight) {\scriptsize\tt \ldots};

\draw[<->] (0,{\colHeight + .2}) -- node [above] {\footnotesize header} (5,{\colHeight + .2});
\draw[<->] (5,{\colHeight + .2}) -- node [above] {\footnotesize block payload} (9,{\colHeight + .2});

\draw[->] (3.5,0) -- (3.5,-.2) -- (5.1,-.2) --  (5.1,-.0);
\draw[->, dashed] (5.5,0) -- (5.5,-.2) -- (9,-.2);

\draw[->] (4.5,0) -- (4.5,-.4) -- (7.1,-.4) --  (7.1,0);
\draw[->, dashed] (7.5,0) -- (7.5,-.4) -- (9,-.4);

\end{tikzpicture}
\caption{Real span layout}
\label{fig:span-header}
\end{figure}

The real-span header layout is shown in Figure~\ref{fig:span-header}. A 
\impl{link} field is used to link up spans when necessary, i.e., it is used to link
spans in the span-pool as well as in the set of reusable spans. The \impl{epoch}
field is used to uniquely identify a span's state within its life cycle (see
below). The local and remote free list contained in a span are encoded in the
fields \impl{local f-list} and \impl{remote f-list}, respectively. A span's
owning LAB is encoded in the \impl{owner} field.

The fields of a LAB are: an \impl{owner} field that uniquely identifies a LAB;
for each size class a field that refers to the hot span, called
\impl{hot\_span}; and per size class the set of reusable spans kept in a field
\impl{reusable\_span}.

\OurSubsubsection{Owner encoding.}
The \impl{owner} field consists of two parts, an actual identifier (16 bits) of
the owning LAB and a reference (48 bits) to the owning LAB. The whole field fits
in a single word and can be updated atomically using compare-and-swap
instructions.  Note that upon termination of the last thread that is assigned to
a LAB, the \impl{owner} is set to \impl{TERMINATED}. Subsequent reuses of the
LAB (upon assigning newly created threads to it) result in a different owner,
i.e., the actual identifier is different while the reference to the LAB stays
the same. Also note that due to thread termination a span's owning LAB might
have a different \impl{owner} than the span's \impl{owner} field indicates.

\OurSubsubsection{Epoch encoding.}
The \impl{epoch} field is a single word that encodes a span's state and an ABA
counter. The states hot, free, reusable, and floating are encoded in the upper
parts (bitwise) of the word. The ABA counter (encoded in the rest of the word)
is needed for versioning spans as the state alone is not enough to uniquely
encode a state throughout a span's life cycle. E.g., one thread can observe a
reusable span that after the last free is empty. Since freeing the object and
transitioning the span into the state is not an atomic operation, another thread
can now observe this span as empty (because it has been delayed after an earlier
free operation) and put it into the span-pool. This span can now be reused by
the same thread in the same size class ultimately ending up in the state
reusable, but not completely empty. At this point the thread that initially
freed the last block in the previous round needs to be prevented from
transitioning it into state free.

\OurSubsubsection{Remote f-list encoding.}
We use a Treiber stack~\cite{Treiber86} to implement the remote free list in a
span. The top pointer of the stack is stored in its own cache line in the span
header. Furthermore, we keep the number of blocks in the stack's \impl{top}
pointer. This number is increased on each put operation. A single call is used
to retrieve all blocks from this free list by atomically setting the stack's
\impl{top} pointer to \impl{NULL} (implicitly setting the block count to 0).
Note that generating a new state (putting and retrieving all blocks) only
requires the \impl{top} pointer. As a result special ABA handling is not needed
(ABA can occur, but is not a problem)~\footnote{For detailed explanations of the
ABA problem see~\cite{Michael:IEEE04}.}.

\begin{figure}[t]
\begin{lstlisting}[
  caption={Auxiliary structures and methods},
  label={lst:lab-aux-methods},  xleftmargin=3.0ex]
// Constant indicating terminated LABs. 
Int TERMINATED;

LAB get_lab(Int owner);                  (*\CL{aux-getlab}*)
Bool is_orphan(Span span);               (*\CL{aux-isorphan}*)

Span { /* Free list implementations omitted. */
  Int epoch;    (*\CL{aux-span-decl-start}*)
  Int owner;    (*\CL{aux-span-decl-end}*)
  Int size_class;

  Bool try_mark_hot(Int old_epoch);     (*\CL{aux-span-states-start}*)
  Bool try_mark_floating(Int old_epoch);
  Bool try_mark_reusable(Int old_epoch);
  Bool try_mark_free(Int old_epoch);    (*\CL{aux-span-states-end}*)

  Bool try_refill_from_remotes();       (*\CL{aux-span-refill}*)
  Bool try_adopt(Int new_owner);        (*\CL{aux-span-adopt}*)
}

Set { /* Set implementation omitted. */
  Int owner;

  void open(Int owner);(*\CL{aux-set-open}*)
  void close();(*\CL{aux-set-close}*)  // Sets owner to TERMINATED;

  Span get();                        (*\CL{aux-set-start-1}*)
  Bool put(Int old_owner, Span span);
  Bool try_remove(Int old_owner, Span span);(*\CL{aux-set-end-1}*)
}
\end{lstlisting}
\end{figure}

\bigbreak
Listing~\ref{lst:lab-aux-methods} provides an overview of auxiliary methods on
spans and sets for reusable spans.

\sloppypar{

Recall that an \impl{owner} field embeds a reference to the corresponding LAB,
which can be retrieved using \impl{get\_lab}~(\lineRef{aux-getlab}).
Furthermore, the function \impl{is\_oprhan}~(\lineRef{aux-isorphan}) is used to
check whether the given span is an orphan, i.e., all threads assigned to its
owning LAB have terminated before all blocks have been returned.

}

A span then contains the previously mentioned \impl{epoch} and \impl{owner}
fields~(\lineRangeRef{aux-span-decl-start}{aux-span-decl-end}). The
methods that try to mark a span as being in a specific
state~(\lineRangeRef{aux-span-states-start}{aux-span-states-end}) all
take an epoch value and try to atomically change it to a new value that has the
corresponding state bits set and the ABA counter increased. These calls are then
used in the actual algorithm for allocation and deallocation to transition a
span from one state into another. The method
\impl{try\_refill\_from\_remotes}~(\lineRef{aux-span-refill}) is used to move
remote blocks (if there are more available then \emph{reusability threshold})
from the remote free list to the local one. The method
\impl{try\_adopt}~(\lineRef{aux-span-adopt}) is used to adopt orphaned
spans, i.e., atomically change their owning LAB.

Maintaining reusable spans should not have a noticeable performance impact
(latency of allocation and deallocation) --- which of course suggests using fast
and scalable rather than non-scalable and slow  data structures. Our design
provides constant time \impl{put}, \impl{get}, and \impl{remove} of arbitrary
spans~(\lineRangeRef{aux-set-start-1}{aux-set-end-1}). Furthermore, reusable
spans are cleaned upon termination of the last thread assigned to a LAB,
requiring \impl{open} and \impl{close}
methods~(\lineRangeRef{aux-set-open}{aux-set-close}) that effectively prohibit
\impl{put} and \impl{remove} methods accessing a set when no owner (i.e.
\impl{TERMINATED}) is present or the owner is different than the one
provided as parameter. For details on thread termination see below. Contention
on sets of reusable spans is low as the sets are segregated by size class and
LABs. For the implementation of reusable sets of spans in scalloc we use a
lock-based deque. We are aware of lock-free implementations of
deques~\cite{Dodds:POPL15} that can be enhanced to be usable in scalloc.
However, the process of cleaning up the set at thread termination (see below)
requires wait-freedom as other threads may still be accessing the data
structure. Helping approaches can be used to (even
efficiently~\cite{Kogan:PPoPP12}) solve this problem. Experiments suggest that
contention on these sets is low and we thus keep the implementation simple.

\begin{figure}[ht!]
\begin{lstlisting}[
  caption={Frontend: Allocation, deallocation, and thread termination and
           initialization},
  label={lst:scalloc-frontend},  xleftmargin=3.0ex]
LAB {
  Span hot_span[NUM_SIZE_CLASSES]; (*\CL{main-decl-start}*)
  Set reuseable_spans[NUM_SIZE_CLASSES];
  Int owner; (*\CL{main-decl-end}*)

  // Retrieve a span from the set of reusable spans
  // or the span-pool.
  Span get_span(Int size_class): (*\CL{main-get-span}*)
    Span span; 
    do: (*\CL{main-getspan-reusable1}*)
      span =  reuseable_spans[size_class].get();
      if span != NULL &&
          span.try_mark_hot(span.epoch):(*\CL{main-getspan-markhot1}*)
        return span;
    until span == NULL; (*\CL{main-getspan-reusable2}*)
    span = span_pool.get(size_class); (*\CL{main-getspan-sp}*)
    span.try_mark_hot(span.epoch);//always succeeds(*\CL{main-getspan-markhot2}*)
    return span;

  Block allocate(Int sc /* size class */): (*\CL{main-alloc}*)
    if hot_span[sc] == NULL:
      hot_span[sc] = get_span(sc); (*\CL{main-alloc-getspan1}*)
    Block block = hot_span[sc].allocate_block(); (*\CL{main-alloc-alloc}*)
    if block == NULL:
      // Case of empty local free list.
      if hot_span[sc].try_refill_from_remotes(): (*\CL{main-alloc-refill}*)
        block = hot_span[sc].allocate_block(); (*\CL{main-alloc-alloc2}*)
        return block;
      hot_span[sc].try_mark_floating(span.epoch); (*\CL{main-alloc-floating}*)
      hot_span[sc] = get_span(sc); (*\CL{main-alloc-new}*)
      block = hot_span[sc].allocate_block(); (*\CL{main-alloc-alloc3}*)
    return block;                                       

  void deallocate(Block block):  (*\CL{main-free}*)
    Span span = span_from_block(block);
    Int sc = span.size_class;
    Int old_owner = span.owner; (*\CL{main-free-span-old-owner}*)
    Int old_epoch  = span.epoch; (*\CL{main-free-span-old-epoch}*)
    span.free(block, owner); (*\CL{main-free-free}*)
    if span.is_orphan(): (*\CL{main-free-orphan1}*)
      span.try_adopt(owner); (*\CL{main-free-orphan2}*)
    if span.free_blocks() > REUSABILITY_THRESHOLD: (*\CL{main-free-reusable-1}*)
      if span.try_mark_reusable(old_epoch): (*\CL{main-free-reusable-2}*)
        old_owner.reuseable_spans[sc].put(
            old_owner, span);(*\CL{main-free-reusable-3}*)
    if span.is_full(): (*\CL{main-free-full-1}*)
      if span.try_mark_free(old_epoch): (*\CL{main-free-full-2}*)
        old_owner.reuseable_spans[sc].try_remove(
            old_owner, span);(*\CL{main-free-full-3}*)
        span_pool.put(span); (*\CL{main-free-full-4}*)

  void terminate(): (*\CL{main-terminate}*)
    for sc in size_classes: (*\CL{main-terminate-sc-start}*)
      reuseable_spans[sc].close(); (*\CL{main-terminate-close}*)
      hot_span[sc].try_mark_floating(
          hot_span[sc].epoch); (*\CL{main-terminate-floating-1}*)
      hot_span[sc] = NULL;
      Span span;
      do:
        span = reuseable_spans[sc].get();
        span.try_mark_floating(span.epoch); (*\CL{main-terminate-floating-2}*)
      until span == NULL; (*\CL{main-terminate-sc-end}*)
    owner = TERMINATED;

  void init(Int new_owner):
    owner = new_owner;
    for sc in size_classes:
      reuseable_spans[sc].open(new_owner); (*\CL{main-init}*)
}
\end{lstlisting}
\end{figure}

\bigbreak
Listing~\ref{lst:scalloc-frontend} illustrates the main parts of scalloc's
frontend. For simplicity we omit error handling, e.g. returning out of memory.
Recall that each LAB is assigned an \impl{owner} and holds for each size class a
reference to the \impl{hot\_span} and the
\impl{reusable\_spans}~(\lineRangeRef{main-decl-start}{main-decl-end}).

The method \impl{get\_span}~(\lineRef{main-get-span}) is used to retrieve new
spans, either from the reusable
spans~(\lineRangeRef{main-getspan-reusable1}{main-getspan-reusable2}), or from
the span-pool~(\lineRef{main-getspan-sp}). The calls to
\impl{try\_mark\_hot} on \lineRef{main-getspan-markhot1} and
\lineRef{main-getspan-markhot2} represent the transitions 
\trans{free}{hot} and \trans{reusable}{hot}, respectively. Note that the
transition \trans{free}{hot} does not compete with any other threads.

The method \impl{allocate}~(\lineRef{main-alloc}) is used to allocate a single
block in a hot span. If no hot span is present a new one is obtained using
\impl{get\_span}~(\lineRef{main-alloc-getspan1}). The hot span is then used to
retrieve a block from the local free list of a
span~(\lineRef{main-alloc-alloc}). If this attempt fails because the local free
list is empty, the remote free list is inspected. If enough (with respect to
\emph{reusability threshold} represented as \impl{REUSABILITY\_THRESHOLD})
remotely freed blocks are available, they are moved to the local free
list~(\lineRef{main-alloc-refill}), just before actually allocating a new
block~(\lineRef{main-alloc-alloc2}). If not enough remotely freed blocks are
available the current hot span is marked as
floating~(\lineRef{main-alloc-floating}), i.e., the hot span takes the
transition
\trans{hot}{floating}, and a new hot span is
retrieved~(\lineRef{main-alloc-new}). The block is then allocated in the new hot
span~(\lineRef{main-alloc-alloc3}).

The method \impl{deallocate}~(\lineRef{main-free}) is used to free a single
block. Since freeing a block and transitioning spans through states are
non-atomic operations, the \impl{owner} and \impl{epoch} values of a span are
stored before freeing the
block~(\lineRangeRef{main-free-span-old-owner}{main-free-span-old-epoch}). The
span's free call~(\lineRef{main-free-free}) then puts the block into the
corresponding free list (local or remote, depending on the owner of the span).
If after this free, the number of free blocks is larger than the
\emph{reusability threshold}~(\lineRef{main-free-reusable-1}), the span is put
into state reusable (\trans{floating}{reusable}). Similar to other state
transitions, this action is serialized through
\impl{try\_mark\_reusable}~(\lineRef{main-free-reusable-2}). The succeeding
thread then also tries to insert the span into the set of reusable spans for
this size class~(\lineRef{main-free-reusable-3}). Note that this call takes the
old owner as parameter to prohibit inserting into a reusable set of a LAB that
has either no owner or an owner that is different from the old owner. Similarly,
making the transition
\trans{reusable}{free} requires marking it as free through
\impl{try\_mark\_free}~(\lineRef{main-free-full-2}). Note that marking a span as
free competes with reusing it in \impl{get\_span}. After successfully
marking it as free the span can be removed (if needed) from the set of reusable
spans~(\lineRef{main-free-full-3}). Finally, the span is put into the
span-pool~(\lineRef{main-free-full-4}).

\OurSubsubsection{Thread Termination.}
Similar to others~\cite{Hudson:ISMM06}, we refer to spans that have not yet been
transitioned into the state free (because they contain live blocks) while all
threads assigned to their owning LAB have terminated as \emph{orphaned} spans.
Since LABs do not necessarily have references to all owned spans (there exist no
references to floating spans) a span cannot be declared as orphaned by setting a
flag. Instead, orphaned spans can be detected by comparing a span's owner against
the owner that is set in the owning LAB. Owner fields that differ or a LAB owner
set to \impl{TERMINATED} indicates an orphaned span. Orphaned spans are always
floating and adopted by threads upon freeing a block in these
spans~(\lineRangeRef{main-free-orphan1}{main-free-orphan2}). LAB cleanup happens
in \impl{terminate}~(\lineRef{main-terminate}) where for each size
class~(\lineRangeRef{main-terminate-sc-start}{main-terminate-sc-end}) the
reusable spans set is closed~(\lineRef{main-terminate-close}), and all spans
(hot and reusable) are transitioned into state
floating~(\lineRef{main-terminate-floating-1} and
\lineRef{main-terminate-floating-2}). For reusable spans this transition
competes with \trans{reusable}{free}~(\lineRef{main-free-full-2}), where a
potential last free of a block in a span triggers putting the span into the
span-pool. Finally, the LAB is marked as terminated and consequently all spans
that are not free can be observed as orphaned. Reusing a LAB later on requires
setting a new unique owner~(\lineRef{main-init}).

\OurSubsubsection{Handling Large Objects.}
Scalloc provides span-based allocation of blocks of size less than or equal to
1MB and relies on conventional \impl{mmap} for all other objects. For allocation
this means that the frontend just forwards the allocation request to an
allocator that just mmaps the block and adds a header containing the required
size. Deallocation requires checking whether a block has been allocated in a
span or not. However, since spans are contained in a single large arena this
check is cheap (xor-ing against aligned arena boundary). Depending on whether
the block has been allocated in a span or not, the request is just forwarded
appropriately.

\OurSubsubsection{Unwritten Rules.}
The illustrated concepts yield a design that provides scalability on a
multi-core system while keeping memory compact with respect to to a reusability threshold.
To this end we would like to note that being competitive in absolute terms
requires an implementation that forces strict inlining of code, careful
layout of thread-local storage, and intercepting thread creation and
termination. Without those techniques absolute performance suffers from
overheads of  function calls as well as cache misses (for unnecessarily checking
conditions related to thread-local storage).

\section{Related Work}\label{sec:related-work}

We first discuss related work on concurrent data structures and then on existing
concurrent memory allocators.

Concurrent data structures in the fast path of the frontend as well as the
backend of scalloc are lock-free~\cite{Herlihy08}. A recent trend towards semantically
relaxed concurrent data structures~\cite{Afek:OPODIS10, Henzinger:POPL13} opens
doors for new design ideas and even greater performance and scalability so that
hierarchies of spans (buffers in general) can be avoided and may be utilized
globally across the whole system. The concurrent data structures in scalloc are
pools, allowing in principle every data structure with pool semantics to be
used. However, unlike segment queues~\cite{Afek:OPODIS10} and $k$-Stacks~\cite{Henzinger:POPL13}, Distributed Queues~\cite{Haas:CF13} with Treiber
stacks~\cite{Treiber86}, as used in scalloc, do not require dynamically
allocated administrative memory (such as sentinel nodes), which is important for
building efficient memory allocators. The data structures for reusable spans within a TLAB 
are implemented using locks but could in principle be replaced with 
wait-free sets, which nowadays can be implemented almost as efficiently
as their lock-free counterparts~\cite{Kogan:PPoPP12}.

Many concepts underlying scalloc such as size classes, hierarchical allocation
(local allocation buffers, spans), and local memory caching in so-called private
heaps (span ownership) have already been introduced and studied in thread-local
variants~\cite{Wilson:ISMM95,Berger:ASPLOS00}. Scalloc borrows from
some of these concepts and integrates them with lock-free
concurrent data structures, and introduces new ideas like virtual spans.

In our experiments we compare scalloc to some of the best and most
popular concurrent memory allocators: Hoard
(git-13c7e75), jemalloc (3.6.0), llalloc (1.4), ptmalloc2\footnote{Since ptmalloc3 performs worse than ptmalloc2, we exclude ptmalloc3
from our experimental evaluation.} (libc 2.19), Streamflow (git-41aa80d),  
SuperMalloc (git-bd7096f), Intel TBB allocator (4.3), and TCMalloc (googleperftools 2.1). 
\mbox{McRT-Malloc}~\cite{Hudson:ISMM06} is left out of our comparison because of a
missing implementation. For Michael's allocator~\cite{Michael:PLDI04} there
exists no reference implementation
--- an implementation for x86-64 by the Streamflow authors crashes for all our
benchmarks; we have received another implementation\footnote{In a private correspondence
we received pointers to the  Amino Concurrent Building Blocks
(\url{http://amino-cbbs.sourceforge.net/}) malloc
implementation.}
which unfortunately does not perform and scale as we expect from the original paper. We thus
decided to leave the Michael allocator out of our comparisons.
ptmalloc2~\cite{Gloger:ptmalloc2} extends Doug Lea's malloc~\cite{Lea:dlmalloc}
(dlmalloc; 2.7.x) and is part of the GNU libc library.
jemalloc~\cite{Evans:jemalloc} is the default allocator in FreeBSD and NetBSD
and has been integrated into products of Mozilla (like Firefox) and Facebook.
TCMalloc~\cite{tcmalloc} is Google's counter-part to jemalloc, also aiming at
performance and scalability. TBB~\cite{tbb} is maintained by Intel as part of
their thread building block library which aims at easy creation of fast and
scalable multi-threaded programs.
llalloc~\cite{llalloc} is an allocator developed by Lockless Inc.
Hoard~\cite{Berger:ASPLOS00} and
Streamflow~\cite{Schneider:ISMM06} are both academic allocators known to perform well.
SuperMalloc~\cite{Kuszmaul:ISMM15} is another recently developed academic allocator introducing, to our knowledge independently, after scalloc's virtual spans, the idea of segmenting virtual address space for
uniform treatment of different-sized objects.
Note that the concept of virtual spans may readily be used in
other allocators and is orthogonal to the actual allocator design.
Moreover, mapping virtual memory in a scalable fashion, e.g. as in
RadixVM~\cite{Clements:EuroSys13}, does not solve the problem of designing a
competitive allocator. Memory fragmentation and system-call overhead still need
to be managed.

All mentioned allocators create private heaps in one way or another. This design
has proven to reduce contention and (partially) avoid false sharing. Scalloc is
no different in this aspect as it also creates private heaps (span ownership)
and exchanges space between threads (through the span-pool).

Another aspect all allocators have in common are heaps segregated by size
classes for spatial locality. It makes object headers obsolete (except for
coalescing which is why ptmalloc2 uses them).

Allocators implementing private heaps without returning remotely freed objects
to the allocating threads suffer from unbounded blowup fragmentation in
producer-consumer workloads~\cite{Berger:ASPLOS00}. Hence, it is necessary to
transfer remotely freed memory back to the heap it was allocated on.

ptmalloc2 solves the blowup fragmentation problem by globally locking and then
deallocating the block where it has been allocated. TCMalloc and jemalloc both
maintain caches of remotely freed objects which are only returned to a global
heap when reaching certain thresholds. Hoard allocates objects in superblocks
which are similar to the spans in scalloc. Unlike scalloc,  Hoard returns
remotely freed objects in a hierarchical fashion, first by deallocating the
objects in the superblocks in which they were allocated, then by transferring
superblocks between private heaps, and eventually by transferring superblocks
from private heaps to a global heap. For this purpose, Hoard locks the involved
superblocks, private heaps, and the global heap. llalloc, Streamflow, and TBB
maintain a private and a public free-list per thread and size class. The public
free lists are implemented using lock-free algorithms. Scalloc does exactly
the same.

To this end we would like to note that the frontends of llalloc and Streamflow are to some extent similar to scalloc's frontend. However, both allocators require
cleaning up empty spans in allocation calls, and use different strategies for
large objects and backends.

Another common practice in many allocators is to handle small and big objects,
whatever the threshold between small and big is, in separate sub-allocators
which are typically based on entirely different data structures and algorithms.
jemalloc, llalloc, ptmalloc2, and TCMalloc are such hybrid allocators,
whereas Hoard, scalloc, Streamflow, SuperMalloc, and TBB are not. ptmalloc2 manages big
objects in best-fit free-lists, jemalloc and TCMalloc round the size of big
objects up to page multiples and allocate them from the global heap, and llalloc
maintains a binary tree of big objects (c.f. binary buddy system) in a
large portion of memory obtained from the operating system. Huge objects, again
whatever the threshold between big and huge is, are handled by the operating
system in all considered allocators including scalloc. The principle challenge
is to determine the thresholds between small and big, for hybrid allocators, as
well as between big and huge, for all allocators. Scalloc addresses that
challenge, through virtual spans, by removing the threshold between small and
big objects and by making the threshold between big and huge so large that it is
likely to be irrelevant in the foreseeable future for most existing
applications.

A concept related to virtual spans called spaces appears in the Memory Management Toolkit (MMTk) for managed languages~\cite{Blackburn:ICSE04}.
We note that the generally poor performance of 
SuperMalloc for concurrent workloads shows that virtual spans alone do not suffice for achieving competitive temporal and spatial performance and scalability. Virtual spans alone only simplify allocator design as they enable uniform treatment of small and big objects, and reduce memory consumption. As shown in our experimental evaluation, the combination of virtual spans
with a scalable backend  and a high-performance frontend is crucial for
achieving competitive performance and scalability.

\section{Experimental Evaluation}\label{sec:experiments}

In our experiments we compare scalloc with other allocators and with other scalloc configurations on synthetic and real workloads.
Our evaluation is structured in two parts.
The first part of the evaluation covers well known allocator workloads from
the literature (threadtest, shbench, larson)~\cite{Berger:ASPLOS00,shbench,
Larson:ISMM98} as well as the single-threaded allocation intensive workload
483.xalancbmk from the SPEC CPU2006 suite~\cite{Henning:2006} that are known
to  perform interesting usage patterns, e.g. threadtest provides a completely
thread-local workload for batched allocation and deallocation of objects.
Furthermore, the evaluation also employs workloads generated with ACDC~\cite{Aigner:ISMM13}, a
benchmarking tool that can be configured to emulate virtually any relevant
workload characteristic not covered by existing benchmarks, e.g. 
producer-consumer patterns, varying object sizes, and different object access patterns.
The second part of the evaluation focuses on our design decisions implemented in
scalloc. In particular, we show the impact of effectively
disabling  key features of scalloc like virtual spans.

We have also experimented with application benchmarks and generally found that scalloc either performs as other competitive allocators
or better.  However, in our and others'~\cite{Michael:PLDI04} experience most concurrent applications either use custom allocation or tailor their behavior so that very few threads allocate concurrently, as is the case in e.g. the Chrome web browser. We see our work as a step towards providing an infrastructure that changes this practice. 

A summary of all
benchmarks can be found in Table~\ref{tbl:benchmarks}. A detailed description
of each experiment is presented in the corresponding subsection. 

All experiments ran on a UMA machine with four 10-core 2GHz Intel Xeon E7-4850
processors supporting two hardware threads per core, 32KB 8-way associative L1
data cache and 32KB 4-way associative L1 instruction cache per core, 256KB
unified 8-way associative L2 cache per core, 24MB unified 24-way associative L3
cache per processor, 128GB of main memory, and Linux kernel version 3.8.0.

Note that recent Linux kernels provide the ability to use transparent huge 
pages\footnote{See
\url{https://www.kernel.org/doc/Documentation/vm/transhuge.txt}} as backing
store for regular pages, i.e., huge pages can be used by the kernel as physical
page frames for regular pages. Since this feature interferes with any mechanism
relying on on-demand paging, e.g. calling \texttt{madvise} to return memory, we have
disabled it in all experiments except one where we use it to
disable virtual spans intentionally. 
Transparent huge pages also impact other allocators, e.g. jemalloc.

There are two configurable parameters in scalloc,
\impl{MADVISE\_THRESHOLD} and \impl{REUSABILITY\_THRESHOLD}. We choose
\impl{MADVISE\_THRESHOLD} to be 32KB which is the smallest real-span
size. In principle one can set the threshold as low as the system page size,
effectively trading performance for lower memory consumption, but as spans are
subject to reuse at all times this is not necessary. Furthermore we choose
\impl{REUSABILITY\_THRESHOLD} to be 80\%, i.e., spans may be reused as soon as
80\% of their blocks are free.
Span reuse is useful in workloads that exhibit irregular
allocation and deallocation patterns.
Since span reuse optimizes memory consumption with negligible overhead
we have also considered a configuration that does not reuse spans
and discuss the results but do not show the data for clarity.

Unless explicitly stated we report the arithmetic mean of 10 sample runs
including the 95\% confidence interval based on the corrected sample standard
deviation. For memory consumption we always report the resident set size (RSS).
The sampling frequency  varies among experiments and has been chosen high enough
to not miss peaks in memory consumption between samples. Since most benchmarks
do not report memory consumption we employ an additional process for measuring
the RSS. As a result, benchmarks like threadtest, larson, and shbench only scale
until 39 threads. We still report the 40 threads ticks to illustrate this
behavior.

\begin{table}[t]
\centering
\caption{Summary of benchmarks}
\label{tbl:benchmarks}
\footnotesize
\begin{tabular}{lcccc}
\toprule
{\sc Benchmark} & {\sc Object Size} & {\sc Local} & {\sc Remote}  & {\sc Thread} \\
& {\sc  [Bytes]} & {\sc Frees} & {\sc Frees} & {\sc Term.} \\
\midrule
\multicolumn{5}{l}{\textsc{Single-threaded}}\\
483.xalancbmk & 1-2M & 100\% & 0\% & no \\
\midrule
\multicolumn{5}{l}{\textsc{Multi-threaded}}\\
Threadtest & 64 & 100\% & 0\% & no \\
Shbench & 1-8 & 100\% & 0\% & no \\
Larson & 7-8 & $\ge 99\%$ & $<1\%$ & yes \\
Prod.-Cons. & 16-512  & 
  \multicolumn{2}{c}{see Section~\ref{sec:acdc-sharing}} & no \\
False Sharing & 8 & 
  \multicolumn{2}{c}{see Section~\ref{sec:false-sharing}} & no \\
Object Sizes & 16-4M & 100\% & 0\% & no \\
Spatial Locality & 16-32 & 100\% & 0\% & no \\
\midrule
\multicolumn{5}{l}{\textsc{Design Decisions}}\\
Virtual Spans & 16-4M & 100\% & 0\% & no\\
Span-Pool & 256 & 100\% & 0\% & no\\
Frontend & 64 & 100\% & 0\% & no\\
\bottomrule
\end{tabular}
\end{table}

\subsection{Single-threaded Workload: 483.xalancbmk}

We compare different allocators on the 483.xalancbmk workload of the SPEC
CPU2006 suite which is known to be an allocation intense single-threaded
workload~\cite{Serebryany:USENIX12}.

Figure~\ref{fig:xalan} reports a benchmark score called ratio where a higher
ratio means a lower benchmark running time. The results omit data for Streamflow
because it crashes. Scalloc (among others) provides a significant improvement
compared to ptmalloc2 but the differences among the best performing allocators
including scalloc are small. Nevertheless, the results demonstrate competitive
single-threaded temporal performance of scalloc. Note that the SPEC suite does
not provide metrics for memory consumption.

\subsection{Thread-local Workloads}\label{sec:thread-local-workloads}

We evaluate the performance of allocators in workloads that only consist of
thread-local allocations and deallocations. Recall that scalloc only allocates
blocks in a hot span of a given size class. Hence, workloads that perform more
consecutive allocations in a single size class than a span can hold blocks (i.e. the
working set is larger than the real-span size) result in benchmarking the span-pool.

\subsubsection*{Threadtest}

Threadtest~\cite{Berger:ASPLOS00} allocates and deallocates objects of the same
size in a configurable number of threads and may perform a configurable amount
of work in between allocations and deallocations. For a variable number of
threads $t$, the benchmark is configured to perform $10^4$ rounds per thread of
allocating and deallocating $\frac{10^5}{t}$ objects of 64 bytes. For temporal
performance we show the speedup with respect to single-threaded ptmalloc2
performance.

\begin{figure}[t]
  \centering
  \resizebox{\columnwidth}{!}{%
    \graphicspath{{./fig/xalancbmk/}}%
    \input{fig/xalancbmk/ratio.tex}%
  }%

  \ourSubcaption{Single-threaded temporal performance: SPEC CPU2006 
              483.xalancbmk}
  \label{fig:xalan}
\end{figure}

In scalloc, objects of 64 bytes are allocated in 64-byte blocks in spans with
a real-span size of 32KB. Since allocations and deallocations are performed in rounds
reusing of spans has no effect on memory consumption. The overhead of adding
and removing spans to the reusable set is negligible.

Figure~\ref{fig:threadtest-performance} illustrates temporal performance where
all allocators but jemalloc scale until 39 threads with only the absolute
performance being different for most allocators. Since the working set for a
single round of a thread is roughly 6.4MB, allocators are forced to interact
with their backend. The results suggest that the span-pool with its strategy of
distributing contention provides the fastest backend of all considered
allocators. The memory consumption shown in Figure~\ref{fig:threadtest-memory}
suggests that threads do not exhibit a lock-step behavior, i.e., they run out of
sync with respect to their local rounds, which ultimately manifests in lower memory
consumption for a larger number of threads. The span-pool supports this behavior
by providing a local fast path with a fall back to scalable global sharing of
spans.

\subsubsection*{Shbench}

Similar to threadtest, shbench~\cite{shbench} exhibits a thread-local behavior
where objects are allocated and deallocated in rounds. Unlike threadtest though,
the lifetime of objects is not exactly one round but varies throughout the
benchmark. For a variable number of threads $t$, the benchmark is configured to
perform $10^6$ rounds per thread of objects between 1 and 8 bytes in size. We
exclude Streamflow from this experiment because it crashes for more than 1
thread in the shbench workload. For temporal performance we show the speedup
with respect to single-threaded ptmalloc2.

Figure~\ref{fig:shbench-performance} shows the performance results. As objects
survive rounds of allocations the contention on the span-pool is not as high as
in threadtest. The memory consumption in Figure~\ref{fig:shbench-memory}
indicates that the absolute performance is determined by span sizes (or
other local buffers). Allocators that keep memory compact in this benchmark also
suffer from degrading absolute performance. Scalloc performs better than all
other allocators except for llalloc which consumes more memory. Note that
reusing spans in this benchmark has an impact on memory consumption. Scalloc is
configured to reuse at 80\% free blocks in a span. Disabling reusing of spans
results for 20 threads in a memory consumption increase of 14.7\% (while having
no noticeable impact on performance). This suggests that reusing spans in
scalloc is important in non-cyclic workloads. Note that the decrease
of memory consumption at 6 threads is a workload artifact. In contrast, the 
peak in memory consumption for llalloc at 20 threads (which is repeatable)
suggests an allocator artifact.

\begin{figure*}[ht!]

  \vspace{-\smallskipamount}%
  \begin{center}%
  \resizebox{\textwidth}{!}{%
    \graphicspath{{./fig/legends/}}%
    \input{fig/legends/general.tex}}%
  \end{center}%
  \vspace{-\smallskipamount}%

\begin{subfigure}[b]{.5\textwidth}
  \centering
  \resizebox{.9\textwidth}{!}{%
    \graphicspath{{./fig/threadtest/}}%
    \input{fig/threadtest/threadtest-performance.tex}%
  }%

  \ourSubcaption{Speedup}
  \label{fig:threadtest-performance}
\end{subfigure}
\begin{subfigure}[b]{.5\textwidth}
  \centering
  \resizebox{.9\textwidth}{!}{%
    \graphicspath{{./fig/threadtest/}}%
    \input{fig/threadtest/threadtest-memory.tex}%
  }%

  \ourSubcaption{Memory consumption}
  \label{fig:threadtest-memory}
\end{subfigure}
\ourCaption{Thread-local workload: Threadtest benchmark}
\label{fig:threadtest}

\begin{subfigure}[b]{.5\textwidth}
  \centering
  \resizebox{.9\textwidth}{!}{%
    \graphicspath{{./fig/shbench/}}%
    \input{fig/shbench/shbench-performance.tex}%
  }%

  \ourSubcaption{Speedup}
  \label{fig:shbench-performance}
\end{subfigure}
\begin{subfigure}[b]{.5\textwidth}
  \centering
  \resizebox{.9\textwidth}{!}{%
    \graphicspath{{./fig/shbench/}}%
    \input{fig/shbench/shbench-memory.tex}%
  }%

  \ourSubcaption{Memory consumption}
  \label{fig:shbench-memory}
\end{subfigure}
\ourCaption{Thread-local workload: Shbench benchmark}
\label{fig:shbench}

\begin{subfigure}[b]{.5\textwidth}
  \centering
  \resizebox{.9\textwidth}{!}{%
    \graphicspath{{./fig/larson/}}%
    \input{fig/larson/larson-performance.tex}%
  }%

  \ourSubcaption{Throughput}
  \label{fig:larson-performance}
\end{subfigure}
\begin{subfigure}[b]{.5\textwidth}
  \centering
  \resizebox{.9\textwidth}{!}{%
    \graphicspath{{./fig/larson/}}%
    \input{fig/larson/larson-memory.tex}%
  }%

  \ourSubcaption{Memory consumption}
  \label{fig:larson-memory}
\end{subfigure}
\ourCaption{Thread-local workload (including thread termination): Larson benchmark}
\label{fig:larson}

\end{figure*}

\begin{figure*}[ht!]

  \vspace{-\smallskipamount}%
  \begin{center}%
  \resizebox{\textwidth}{!}{%
    \graphicspath{{./fig/legends/}}%
    \input{fig/legends/general.tex}}%
  \end{center}%
  \vspace{-\smallskipamount}%

\begin{subfigure}[b]{.5\textwidth}
  \centering
  \resizebox{.9\textwidth}{!}{%
    \graphicspath{{./fig/prod-cons/}}%
    \input{fig/prod-cons/combined.tex}%
  }%

  \ourSubcaption{Total per-thread allocator time}
  \label{fig:prod-cons-tat}
\end{subfigure}
\begin{subfigure}[b]{.5\textwidth}
  \centering
  \resizebox{.9\textwidth}{!}{%
    \graphicspath{{./fig/prod-cons/}}%
    \input{fig/prod-cons/memcons.tex}%
  }%

  \ourSubcaption{Average per-thread memory consumption}
  \label{fig:prod-cons-mem}
\end{subfigure}
\ourCaption{Temporal and spatial performance for the producer-consumer
            experiment}

\begin{subfigure}[b]{.5\textwidth}%
  \centering
  \resizebox{.9\textwidth}{!}{%
    \graphicspath{{./fig/virtual-spans-40threads/}}%
    \input{fig/virtual-spans-40threads/combined.tex}%
  }%

  \ourSubcaption{Total allocator time}
  \label{fig:objectsize-tat}
\end{subfigure}
\begin{subfigure}[b]{.5\textwidth}
  \centering
  \resizebox{.9\textwidth}{!}{%
    \graphicspath{{./fig/virtual-spans-40threads/}}%
    \input{fig/virtual-spans-40threads/memcons.tex}%
  }%

  \ourSubcaption{Average memory consumption}
  \label{fig:objectsize-mem}
\end{subfigure}
\ourCaption{Temporal and spatial performance for the object-size robustness
experiment at 40 threads}

\end{figure*}

\subsubsection*{Larson (Thread Termination)}

The larson benchmark~\cite{Larson:ISMM98} simulates a multi-threaded server
application responding to client requests. A thread in larson receives a set of
objects, randomly performs deallocations and allocations on this set for a given
number of rounds, then passes the set of objects on to the next thread, and
finally terminates. The benchmark captures robustness of allocators for unusual
allocation patterns including terminating threads. Unlike results reported
elsewhere~\cite{Schneider:ISMM06} we do not observe a ratio of 100\% remote
deallocations as larson also exhibits thread-local allocation and deallocation
in rounds. For a variable number of threads $t$, the benchmark is configured to
last 10 seconds, for objects of 7 bytes (smallest size class for all
allocators), with $10^3$  objects per thread, and $10^4$ rounds per slot of
memory per thread. Unlike the other experiments, the larson benchmark runs for a
fixed amount of time and provides a throughput metric of memory management
operations per second. We exclude Streamflow from this experiment as it
crashes under the larson workload.

Figure~\ref{fig:larson-performance} illustrates temporal performance where all
considered allocators scale but provide different speedups. Similar to shbench,
the rate at which spans get empty varies. The  memory consumption in
Figure~\ref{fig:larson-memory} illustrates that better absolute performance
comes at the expense of increased memory consumption. Furthermore, terminating
threads impose the challenge of reassigning spans in scalloc (and likely also
impose a challenge in other allocators that rely on thread-local data
structures). Similar to shbench, reusing spans before they get empty results in
better memory consumption. Disabling span reuse increases memory consumption by
7.6\% at 20 threads while reducing performance by 2.7\%.

\subsection{Producer-Consumer Workload}\label{sec:acdc-sharing}

This experiment evaluates the temporal and spatial performance of a
producer-consumer workload to study the cost of remote frees and possible blowup
fragmentation for an increasing number of producers and consumers. For that
purpose we configure ACDC such that each thread shares all allocated objects
with all other threads, accesses all local and shared objects, and eventually
the last (arbitrary) thread accessing an object frees it. The probability of a
remote free in the presence of $n$ threads is therefore  $1-1/n$, e.g. running
two threads causes on average 50\% remote frees and  running 40 threads causes
on average 97.5\% remote frees.

Figure~\ref{fig:prod-cons-tat} presents the total time each thread spends in the
allocator for an increasing number of producers/consumers. Up to 30 threads
scalloc and Streamflow provide the best temporal performance and for more than
30 threads scalloc outperforms all other allocators.

The average per-thread memory consumption illustrated in
Figure~\ref{fig:prod-cons-mem} suggests that all allocators deal with blowup
fragmentation, i.e., we do not observe unbounded growth in memory consumption.
However, the absolute differences among different allocators are significant.
Scalloc provides competitive spatial performance where only jemalloc and
ptmalloc2 require less memory at the expense of higher total per-thread
allocator time.

This experiment demonstrates that the approach of scalloc to distributing
contention across spans with one remote free list per span works well in a
producer-consumer workload and that using a lock-based implementation for
reusing spans is not a performance bottleneck.

\subsection{Robustness against False Sharing} 
\label{sec:false-sharing}

False sharing occurs when objects that are allocated in the same cache line are
read from and written to by different threads. In cache coherent
systems this scenario can lead to performance degradation as all caches need to
be kept consistent. An allocator is prone to active false
sharing~\cite{Berger:ASPLOS00} if objects that are allocated by different
threads (without communication) end up in the same cache line. It is prone to
passive false sharing~\cite{Berger:ASPLOS00} if objects that are remotely
deallocated by a thread are immediately usable for allocation by this thread
again.

We have conducted the false sharing
avoidance evaluation benchmark from Berger et. al.~\cite{Berger:ASPLOS00} (including active-false and passive-false benchmarks) to validate
scalloc's design. 
The results we have obtained suggest that most allocators avoid active and passive false sharing. However, SuperMalloc and TCMalloc suffer from both active and passive false sharing,
whereas Hoard is prone to passive false sharing only. 
We omit the graphs because they only show binary results (either false sharing occurs or not).
Scalloc's design ensures that in the cases covered by the active-false and passive-false benchmarks no false sharing
appears, as spans need to be freed to be reused by other threads for
allocation. Only in case of thread termination (not covered by the active-false
and passive-false benchmarks) threads may adopt spans in which other threads still have
blocks, potentially causing false sharing. We have not
encountered false sharing with scalloc in any of our experiments.

\subsection{Robustness for Varying Object Sizes}\label{sec:object-sizes}

We configure the ACDC allocator benchmark~\cite{Aigner:ISMM13} to allocate,
access, and deallocate increasingly larger thread-local objects in 40 threads
(number of native cores) to study the scalability of virtual spans and the span
pool.

Figure~\ref{fig:objectsize-tat} shows the total time spent in the allocator,
i.e., the time spent in malloc  and free. The x-axis refers to intervals $[2^x,
2^{x+2})$ of object sizes in bytes with $4\leq x\leq 20$ at increments of two.
For each object size interval ACDC allocates $2^x$KB of new objects, accesses
the objects, and then deallocates previously allocated objects. This cycle is
repeated 30 times. For object sizes smaller than 1MB scalloc outperforms all
other allocators because virtual spans enable scalloc to rely on efficient
size-class allocation. The only possible bottleneck in this case is accessing
the span-pool. However, even in the presence of 40 threads we do not observe
contention on the span-pool. For objects larger than 1MB scalloc relies on mmap
which adds system call latency to allocation and deallocation operations and is
also known to be a scalability bottleneck~\cite{Clements:EuroSys13}.

The average memory consumption (illustrated in  Figure~\ref{fig:objectsize-mem})
of scalloc allocating small objects is  higher  (yet still competitive) because
the real-spans for size-classes smaller than 32KB have the same size and  {\tt
madvise} is not enabled for them. For larger object sizes scalloc causes the  smallest
memory overhead comparable to jemalloc and ptmalloc2.

This experiment demonstrates the advantages of trading virtual address space
fragmentation for high throughput and low physical memory fragmentation.

\begin{figure}[t!]
  \centering
  \resizebox{\columnwidth}{!}{%
    \graphicspath{{./fig/locality/}}%
    \input{fig/locality/access.tex}%
  }%

  \ourSubcaption{Memory access time for the locality experiment}
  \label{fig:locality-access}
\end{figure}

\begin{figure*}[ht!]

  \vspace{-\smallskipamount}%
  \begin{center}%
  \resizebox{\textwidth}{!}{%
    \graphicspath{{./fig/legends/}}%
    \input{fig/legends/virtual-span-evaluation.tex}}%
  \end{center}%
  \vspace{-\smallskipamount}%

\begin{subfigure}[b]{.5\textwidth}
  \centering
  \resizebox{.9\textwidth}{!}{%
    \graphicspath{{./fig/virtual-spans-evaluation/}}%
    \input{fig/virtual-spans-evaluation/combined.tex}%
  }%

  \ourSubcaption{Total allocator time}
  \label{fig:virtual-spans-tat}
\end{subfigure}
\begin{subfigure}[b]{.5\textwidth}
  \centering
  \resizebox{.9\textwidth}{!}{%
    \graphicspath{{./fig/virtual-spans-evaluation/}}%
    \input{fig/virtual-spans-evaluation/memcons.tex}%
  }%

  \ourSubcaption{Average memory consumption}
  \label{fig:virtual-spans-mem}
\end{subfigure}
\ourCaption{Temporal and spatial performance for the virtual span evaluation}
\end{figure*}

\subsection{Spatial Locality}

In order to expose differences in spatial locality, we configure ACDC to access
allocated objects (between 16 and 32 bytes) increasingly in allocation order
(rather than out-of-allocation order). For this purpose, ACDC organizes
allocated objects either in trees (in depth-first, left-to-right order,
representing out-of-allocation order) or in lists (representing allocation
order). ACDC then accesses the objects from the tree in depth-first,
right-to-left order and from the list in FIFO order.  We measure the total
memory access time for an increasing ratio of lists, starting at 0\% (only
trees), going up to 100\% (only lists), as an indicator of spatial locality.
ACDC provides a simple mutator-aware allocator called compact to serve as
optimal (yet without further knowledge of mutator behavior unreachable)
baseline. Compact stores the lists and trees of allocated objects without space
overhead in contiguous memory for optimal locality.

Figure~\ref{fig:locality-access} shows the total memory access time for an
increasing ratio of object accesses in allocation order.  Only jemalloc and
llalloc provide a memory layout that can be accessed slightly  faster than the
memory layout provided by scalloc. Note that scalloc does not require object headers and
reinitializes span free-lists upon retrieval from the span-pool. For a larger
ratio of object accesses in allocation order, the other allocators improve as well but not as much as llalloc, scalloc, Streamflow, and TBB which approach the
memory access performance of the compact baseline allocator. Note also that we can
improve memory access time with scalloc even more by setting its reusability threshold to 100\%. In this case spans are only reused once they get completely empty and
reinitialized through the span-pool at the expense of higher memory consumption.
We omit this data for consistency reasons.

To explain the differences in memory access time we pick the data points for
ptmalloc2 and scalloc at x=60\% where the difference in memory access time is
most significant and compare the number of all hardware events obtainable using
perf\footnote{See \url{https://perf.wiki.kernel.org}}. While most numbers are similar we
identify  two events where the numbers differ significantly. First, the L1 cache
miss  rate with ptmalloc2 is 20.8\% while scalloc causes a L1 miss rate of
15.2\% at almost the same number of total L1 cache loads. We account this
behavior to more effective cache line prefetching because we observe 272.2M L1
cache prefetches with scalloc and only 119.8M L1 cache prefetches with
ptmalloc2. Second, and related to the L1 miss rate, we observe 408.5M last-level
cache loads with ptmalloc2 but only 190.2M last-level cache loads with scalloc
because every L1 cache miss causes a cache load on the next level. The
last-level cache miss rate is negligible with both allocators suggesting that
the working set (by design) fits into the last-level cache. Note that
last-level cache events in perf include L2 and L3 cache events on our hardware.

\subsection{Design Decisions}\label{sec:scallocs-contributions}

We now evaluate scalloc's design and how the three main contributions, virtual
spans, the scalable backend, and the constant-time frontend influence temporal
and spatial performance. To do so, we provide several different configurations
of scalloc and the benchmarking environment and compare them against each
other in isolated settings in best-effort manner. By best-effort we mean that 
we design our experiments so that they highlight the contribution of an isolated component to
overall performance. Still, frontened evaluations will also include the backend and vice versa, and disabling virtual spans 
is implicit via disabling a kernel feature. 

\subsubsection*{Virtual Spans}

Virtual spans enable uniform treatment of objects across a large range
of different sizes while avoiding physical memory fragmentation through
on-demand paging and explicit \impl{madvise} calls.
The following experiment aims at illustrating
the benefits, cost, and limitations of virtual spans and
ultimately the role played by the paging mechanism of the operating system.

We compare four scalloc configurations on the same workload as in
Section~\ref{sec:object-sizes}, see Figure~\ref{fig:objectsize-tat} and Figure~\ref{fig:objectsize-mem}. The first configuration, scalloc, is the
default configuration used in all previous experiments.  The second
configuration, scalloc-no-madvise, only differs in the policy for returning
memory to the operating system by disabling all \impl{madvise} calls.  Both
configurations are executed with transparent huge pages disabled in the Linux
kernel, i.e., virtual memory is mapped to 4KB pages only.  The third and fourth
configuration are the same as scalloc and scalloc-no-madvise but with
transparent huge pages enabled, i.e., the kernel may switch from 4KB pages to
2MB pages, effectively disabling the advantages of virtual spans.  Therefore
we suffix those configurations with no-virtual-spans.

Figure~\ref{fig:virtual-spans-tat} shows the total allocator time for an
increasing range of object sizes.
The difference between scalloc and  scalloc-no-madvise
is explained by the cost and consequences of calling madvise which
causes consecutive page faults requiring the operating system to zero pages.
However, the memory consumption of scalloc, illustrated in 
Figure~\ref{fig:virtual-spans-mem}, improves almost proportionaly to the cost 
of calling madvise. For small objects up to 256~bytes there are no 
differences because madvise is only called on spans of larger size classes. 
For huge objects (larger than 1MB)
there are no differences because scalloc allocates 
huge objects directly through \impl{mmap}.

Enabling transparent huge pages (for scalloc-no-virtual-spans and scalloc-no-madvise-no-virtual-spans) allows the kernel to allocate 2MB pages
instead of 4KB pages. Since the cost of zeroing 2MB pages is significantly
higher than that of 4KB pages the cost of allocating spans goes up,
especially for small size classes (less than 256~bytes)
where the real span size is small compared
to the virtual span size. Figure~\ref{fig:virtual-spans-tat} shows that
virtual spans perform better with 4KB pages than with 2MB pages. The
memory consumption in Figure~\ref{fig:virtual-spans-mem} increases
dramatically for small size classes because the whole virtual span size is
materialized (but still unused) in physical memory. For larger objects,
calling \impl{madvise} causes the kernel to fall back from 2MB pages to 4KB
pages. As a consequence, the scalloc-no-virtual-spans configuration approaches
the small memory footprint of scalloc. Still the cost of zeroing 2MB pages
and eventually falling back to 4KB pages causes a significant temporal
overhead compared to scalloc. Disabling the advantages of virtual spans by
enabling transparent huge pages and also disabling \impl{madvise} calls (scalloc-no-madvise-no-virtual-spans) causes the highest memory consumption because unused space
in virtual spans is materialized in physical memory and because the kernel
cannot fall back to 4KB pages. The total allocator time is lower compared to
scalloc-no-virtual-spans because no \impl{madvise} calls are necessary.

From the results of this experiment we conclude that the paging mechanism  of
Linux is sufficient for implementing virtual spans efficiently if paging
happens on a 4KB granularity. We also conclude that calling  \impl{madvise}
can trade speed for memory consumption and choosing either setting depends on
the application of scalloc. However, improving the performance of
\impl{madvise} on operating system level would improve scalloc's virtual span design
even further.

\subsubsection*{Scalable Backend}

A consequence of the design around virtual and real spans is that, depending
on the workload and in particular on object sizes, the backend may be subject to
high levels of contention. The following experiment evaluates the design of
the span-pool in terms of performance and scalability, i.e., it shows that
multiple stacks per size class are necessary to deal  with workloads that
cause high contention on the backend. Note that for workloads that only
utilize small objects,  e.g. 8-byte objects, real spans provide enough buffering only
requiring a single stack per size class in the span-pool. We omit separate
plots for this data. However, for larger objects, real spans only provide
a limited amount of blocks requiring a fast and scalable backend. For example, for
256-byte objects the real span size of 32KB amounts to 127 blocks, potentially
resulting in frequent interaction with the span-pool.

The experiment is based on threadtest and configured as in
Section~\ref{sec:thread-local-workloads} with the only difference being that the
benchmark is configured to allocate and deallocate 256-byte objects. We compare
our default scalloc version,
which is based on a span-pool with as many stacks per size
class as threads, with a version that only uses a single stack per size
class, called scalloc-no-span-pool.

Figure~\ref{fig:span-pool-eval-temporal} illustrates that for an increasing
number of threads, a single stack is unable to deal with contention on the
span-pool, effectively resulting in degraded performance. However, in terms of
memory consumption, Figure~\ref{fig:span-pool-eval-memcons} shows that using a
single stack results in better memory utilization as threads synchronize on
a single source of empty spans. Scalloc's design trades the better performance (up to
2.7x) for slightly worse memory utilization (up to 15\%).

\begin{figure*}[ht!]

  \vspace{-\smallskipamount}%
  \begin{center}%
  \resizebox{\textwidth}{!}{%
    \graphicspath{{./fig/legends/}}%
    \input{fig/legends/span-pool-evaluation.tex}}%
  \end{center}%
  \vspace{-\smallskipamount}%

\begin{subfigure}[b]{.5\textwidth}
  \centering
  \resizebox{.9\textwidth}{!}{%
    \graphicspath{{./fig/span-pool-evaluation/}}%
    \input{fig/span-pool-evaluation/threadtest-performance.tex}%
  }%

  \ourSubcaption{Speedup}
  \label{fig:span-pool-eval-temporal}
\end{subfigure}
\begin{subfigure}[b]{.5\textwidth}
  \centering
  \resizebox{.9\textwidth}{!}{%
    \graphicspath{{./fig/span-pool-evaluation/}}%
    \input{fig/span-pool-evaluation/threadtest-memory.tex}%
  }%

  \ourSubcaption{Memory consumption}
  \label{fig:span-pool-eval-memcons}
\end{subfigure}
\ourCaption{Span-Pool evaluation: Threadtest benchmark using 256-byte objects}

  \vspace{-\smallskipamount}%
  \begin{center}%
  \resizebox{\textwidth}{!}{%
    \graphicspath{{./fig/legends/}}%
    \input{fig/legends/frontend-evaluation.tex}}%
  \end{center}%
  \vspace{-\smallskipamount}%

\begin{subfigure}[b]{.5\textwidth}
  \centering
  \resizebox{.9\textwidth}{!}{%
    \graphicspath{{./fig/frontend-evaluation/}}%
    \input{fig/frontend-evaluation/threadtest-performance.tex}%
  }%

  \ourSubcaption{Speedup}
  \label{fig:frontend-eval-temporal}
\end{subfigure}
\begin{subfigure}[b]{.5\textwidth}
  \centering
  \resizebox{.9\textwidth}{!}{%
    \graphicspath{{./fig/frontend-evaluation/}}%
    \input{fig/frontend-evaluation/threadtest-memory.tex}%
  }%

  \ourSubcaption{Memory consumption}
  \label{fig:frontend-eval-memcons}
\end{subfigure}
\ourCaption{Frontend evaluation: Threadtest benchmark using 64-byte objects}

\end{figure*}

\subsubsection*{Constant-time Frontend}

In a similar spirit, we evaluate the performance impact of a constant-time frontend
that returns empty spans to the backend as soon as possible, i.e., upon freeing
the last object contained in a span.

The experiment is based on threadtest and configured as in 
Section~\ref{sec:thread-local-workloads}, see Figure~\ref{fig:threadtest-performance} and
Figure~\ref{fig:threadtest-memory}. As the experiment is configured for small
object sizes, returning memory to the operating system only plays a minor
role. We compare our default scalloc version with a scalloc version that
returns empty spans to the backend at a later point in time in its allocation
slow path, called \mbox{scalloc-reclaim-span-in-allocation}. Note that
threadtest provides no information about per-operation latency and hence the
experiment only shows throughput and scalability, but since each thread
implements a closed system, throughput is indirectly proportional to latency.

Figure~\ref{fig:frontend-eval-temporal} illustrates performance and
scalability.  Both scalloc versions provide almost the same speedup
but differ in memory consumption illustrated in 
Figure~\ref{fig:frontend-eval-memcons}. 
The results indicate that eagerly returning spans as soon as they get
empty rather than delaying reclamation until the next (slow path) allocation
increases the potential for reusing spans by other threads effectively
reducing memory consumption.  At 39 threads the difference in memory
consumption is about 25\% between the two scalloc versions.

\section{Conclusion}

We presented three contributions: (a) virtual spans that enable uniform
treatment of small and big objects; (b) a fast and scalable backend leveraging
newly developed global data structures; and (c) a constant-time (modulo
synchronization) frontend that eagerly returns empty spans to the backend. Our
experiments show that scalloc is either better (threadtest, object sizes,
producer-consumer) or competitive (shbench, larson, mutator locality, SPEC) in
performance and memory consumption compared to other allocators.

To conclude, the problem of high-performance and scalable memory allocation is complex. There may be different solutions. Our solution was guided by an initial idea to design an allocator whose scalability benefits from the scalability of recently developed concurrent data structures. In order to make maximal use of global data structures, we developed virtual spans and additionally the constant-time frontend. It may be interesting to study other applications of global scalable concurrent data structures.

Other interesting future work may be
(1) integrating virtual spans and virtual memory management,
(2) NUMA-aware mapping of real spans, and in particular
(3) dynamically resizing real spans to trade off
LAB provisioning and performance based on run-time feedback
from mutators, similar in spirit to just-in-time optimizations
in virtual machines.

\section*{Acknowledgements}

This work has been supported by the National Research Network RiSE on Rigorous
Systems Engineering (Austrian Science Fund (FWF): S11404-N23,  S11411-N23) and
a Google PhD Fellowship.

\bibliographystyle{abbrvnat}
\bibliography{paper}

\begin{thebibliography}{30}
\providecommand{\natexlab}[1]{#1}
\providecommand{\url}[1]{\texttt{#1}}
\expandafter\ifx\csname urlstyle\endcsname\relax
  \providecommand{\doi}[1]{doi: #1}\else
  \providecommand{\doi}{doi: \begingroup \urlstyle{rm}\Url}\fi

\bibitem[Afek et~al.(2010)Afek, Korland, and Yanovsky]{Afek:OPODIS10}
Y.~Afek, G.~Korland, and E.~Yanovsky.
\newblock Quasi-linearizability: Relaxed consistency for improved concurrency.
\newblock In \emph{Proc. Conference on Principles of Distributed Systems
  (OPODIS)}, pages 395--410. Springer, 2010.
\newblock \doi{10.1007/978-3-642-17653-1_29}.

\bibitem[Aigner and Kirsch(2013)]{Aigner:ISMM13}
M.~Aigner and C.~Kirsch.
\newblock {ACDC}: Towards a universal mutator for benchmarking heap management
  systems.
\newblock In \emph{Proc. International Symposium on Memory Management (ISMM)},
  pages 75--84. ACM, 2013.
\newblock \doi{10.1145/2464157.2464161}.

\bibitem[Berger et~al.(2000)Berger, McKinley, Blumofe, and
  Wilson]{Berger:ASPLOS00}
E.~Berger, K.~McKinley, R.~Blumofe, and P.~Wilson.
\newblock Hoard: a scalable memory allocator for multithreaded applications.
\newblock In \emph{Proc. International Conference on Architectural Support for
  Programming Languages and Operating Systems (ASPLOS)}, pages 117--128. ACM,
  2000.
\newblock \doi{10.1145/384264.379232}.

\bibitem[Berger et~al.(2002)Berger, Zorn, and McKinley]{Berger:OOPSLA02}
E.~Berger, B.~Zorn, and K.~McKinley.
\newblock Reconsidering custom memory allocation.
\newblock In \emph{Proc. Conference on Object-oriented Programming, Systems,
  Languages, and Applications (OOPSLA)}, pages 1--12. ACM, 2002.
\newblock \doi{10.1145/582419.582421}.

\bibitem[Blackburn et~al.(2004)Blackburn, Cheng, and
  McKinley]{Blackburn:ICSE04}
S.~Blackburn, P.~Cheng, and K.~McKinley.
\newblock Oil and water? {High} performance garbage collection in {Java} with
  {MMTk}.
\newblock In \emph{Proc. International Conference on Software Engineering
  (ICSE)}. IEEE, 2004.
\newblock \doi{10.1109/ICSE.2004.1317436}.

\bibitem[Clements et~al.(2013)Clements, Kaashoek, and
  Zeldovich]{Clements:EuroSys13}
A.~Clements, M.~Kaashoek, and N.~Zeldovich.
\newblock {RadixVM}: Scalable address spaces for multithreaded applications.
\newblock In \emph{Proc. ACM European Conference on Computer Systems
  (EuroSys)}, pages 211--224. ACM, 2013.
\newblock \doi{10.1145/2465351.2465373}.

\bibitem[Dodds et~al.(2015)Dodds, Haas, and Kirsch]{Dodds:POPL15}
M.~Dodds, A.~Haas, and C.~Kirsch.
\newblock A scalable, correct time-stamped stack.
\newblock In \emph{Proc. Symposium on Principles of Programming Languages
  (POPL)}, pages 233--246. ACM, 2015.
\newblock \doi{10.1145/2775051.2676963}.

\bibitem[Evans(2006)]{Evans:jemalloc}
J.~Evans.
\newblock A scalable concurrent malloc(3) implementation for freebsd.
\newblock In \emph{Proc. BSDCan}, 2006.

\bibitem[Gloger()]{Gloger:ptmalloc2}
W.~Gloger.
\newblock ptmalloc2 -- a multi-thread malloc implementation.
\newblock \url{http://malloc.de/en/}.

\bibitem[{Google Inc.}()]{tcmalloc}
{Google Inc.}
\newblock gperftools: Fast, multi-threaded malloc() and nifty performance
  analysis tools.
\newblock \url{http://code.google.com/p/gperftools/}.

\bibitem[Haas et~al.(2013)Haas, Henzinger, Kirsch, Lippautz, Payer, Sezgin, and
  Sokolova]{Haas:CF13}
A.~Haas, T.~Henzinger, C.~Kirsch, M.~Lippautz, H.~Payer, A.~Sezgin, and
  A.~Sokolova.
\newblock Distributed queues in shared memory---multicore performance and
  scalability through quantitative relaxation.
\newblock In \emph{Proc. International Conference on Computing Frontiers (CF)},
  pages 17:1--17:9. ACM, 2013.
\newblock \doi{10.1145/2482767.2482789}.

\bibitem[Haas et~al.(2015)Haas, Henzinger, Holzer, Kirsch, Lippautz, Payer,
  Sezgin, Sokolova, and Veith]{Haas:LocLin15}
A.~Haas, T.~Henzinger, A.~Holzer, C.~Kirsch, M.~Lippautz, H.~Payer, A.~Sezgin,
  A.~Sokolova, and H.~Veith.
\newblock Local linearizability.
\newblock \emph{CoRR}, abs/1502.07118, 2015.

\bibitem[Hendler et~al.(2010)Hendler, Incze, Shavit, and
  Tzafrir]{Hendler:SPAA10}
D.~Hendler, I.~Incze, N.~Shavit, and M.~Tzafrir.
\newblock Flat combining and the synchronization-parallelism tradeoff.
\newblock In \emph{Proc. Symposium on Parallel Algorithms and Architectures
  (SPAA)}, pages 355--364. ACM, 2010.
\newblock \doi{10.1145/1810479.1810540}.

\bibitem[Henning(2006)]{Henning:2006}
J.~L. Henning.
\newblock Spec cpu2006 benchmark descriptions.
\newblock \emph{SIGARCH Comput. Archit. News}, 34\penalty0 (4):\penalty0 1--17,
  2006.
\newblock \doi{10.1145/1186736.1186737}.

\bibitem[Henzinger et~al.(2013)Henzinger, Kirsch, Payer, Sezgin, and
  Sokolova]{Henzinger:POPL13}
T.~Henzinger, C.~Kirsch, H.~Payer, A.~Sezgin, and A.~Sokolova.
\newblock Quantitative relaxation of concurrent data structures.
\newblock In \emph{Proc. Symposium on Principles of Programming Languages
  (POPL)}, pages 317--328. ACM, 2013.
\newblock \doi{10.1145/2429069.2429109}.

\bibitem[Herlihy and Shavit(2008)]{Herlihy08}
M.~Herlihy and N.~Shavit.
\newblock \emph{The Art of Multiprocessor Programming}.
\newblock Morgan Kaufmann Publishers Inc., 2008.

\bibitem[Hudson et~al.(2006)Hudson, Saha, Adl-Tabatabai, and
  Hertzberg]{Hudson:ISMM06}
R.~Hudson, B.~Saha, A.-R. Adl-Tabatabai, and B.~Hertzberg.
\newblock Mcrt-malloc: a scalable transactional memory allocator.
\newblock In \emph{Proc. International Symposium on Memory Management (ISMM)},
  pages 74--83. ACM, 2006.
\newblock \doi{10.1145/1133956.1133967}.

\bibitem[{Intel Corporation}()]{tbb}
{Intel Corporation}.
\newblock Thread building blocks (tbb).
\newblock \url{http://threadingbuildingblocks.org}.

\bibitem[Kogan and Petrank(2012)]{Kogan:PPoPP12}
A.~Kogan and E.~Petrank.
\newblock A methodology for creating fast wait-free data structures.
\newblock In \emph{Proc. Symposium on Principles and Practice of Parallel
  Programming (PPoPP)}, pages 141--150. ACM, 2012.
\newblock \doi{10.1145/2145816.2145835}.

\bibitem[Kuszmaul(2015)]{Kuszmaul:ISMM15}
B.~Kuszmaul.
\newblock Supermalloc: A super fast multithreaded malloc for 64-bit machines.
\newblock In \emph{Proc. International Symposium on Memory Management (ISMM)},
  pages 41--55. ACM, 2015.
\newblock \doi{10.1145/2754169.2754178}.

\bibitem[Larson and Krishnan(1998)]{Larson:ISMM98}
P.-A. Larson and M.~Krishnan.
\newblock Memory allocation for long-running server applications.
\newblock In \emph{Proc. International Symposium on Memory Management (ISMM)},
  pages 176--185. ACM, 1998.
\newblock \doi{10.1145/286860.286880}.

\bibitem[Lea()]{Lea:dlmalloc}
D.~Lea.
\newblock A memory allocator.
\newblock \url{http://g.oswego.edu/dl/html/malloc.html}.

\bibitem[{Lockless Inc.}()]{llalloc}
{Lockless Inc.}
\newblock llalloc: Lockless memory allocator.
\newblock \url{http://locklessinc.com/}.

\bibitem[Michael(2004{\natexlab{a}})]{Michael:IEEE04}
M.~Michael.
\newblock Hazard pointers: Safe memory reclamation for lock-free objects.
\newblock \emph{IEEE Trans. Parallel Distrib. Syst.}, 15\penalty0 (6):\penalty0
  491--504, 2004{\natexlab{a}}.
\newblock \doi{10.1109/TPDS.2004.8}.

\bibitem[Michael(2004{\natexlab{b}})]{Michael:PLDI04}
M.~Michael.
\newblock Scalable lock-free dynamic memory allocation.
\newblock In \emph{Proc. Conference on Programming Language Design and
  Implementation (PLDI)}, pages 35--46. ACM, 2004{\natexlab{b}}.
\newblock \doi{10.1145/996893.996848}.

\bibitem[{MicroQuill Inc.}()]{shbench}
{MicroQuill Inc.}
\newblock shbench.
\newblock \url{http://www.microquill.com/}.

\bibitem[Schneider et~al.(2006)Schneider, Antonopoulos, and
  Nikolopoulos]{Schneider:ISMM06}
S.~Schneider, C.~Antonopoulos, and D.~Nikolopoulos.
\newblock Scalable locality-conscious multithreaded memory allocation.
\newblock In \emph{Proc. International Symposium on Memory Management (ISMM)},
  pages 84--94. ACM, 2006.
\newblock \doi{10.1145/1133956.1133968}.

\bibitem[Serebryany et~al.(2012)Serebryany, Bruening, Potapenko, and
  Vyukov]{Serebryany:USENIX12}
K.~Serebryany, D.~Bruening, A.~Potapenko, and D.~Vyukov.
\newblock Addresssanitizer: A fast address sanity checker.
\newblock In \emph{Proc. USENIX Conference on Annual Technical Conference
  (USENIX ATC)}, pages 28--28. USENIX Association, 2012.

\bibitem[Treiber(1986)]{Treiber86}
R.~Treiber.
\newblock Systems programming: Coping with parallelism.
\newblock Technical Report RJ-5118, IBM Research Center, 1986.

\bibitem[Wilson et~al.(1995)Wilson, Johnstone, Neely, and Boles]{Wilson:ISMM95}
P.~Wilson, M.~Johnstone, M.~Neely, and D.~Boles.
\newblock Dynamic storage allocation: A survey and critical review.
\newblock In \emph{Proc. International Workshop on Memory Management (IWMM)},
  pages 1--116. Springer, 1995.
\newblock \doi{10.1007/3-540-60368-9_19}.

\end{thebibliography}

\end{document}